\documentclass{aa}  

\usepackage{graphicx,txfonts,lscape,natbib}
\bibpunct{(}{)}{;}{a}{}{,} 

\begin{document}
   \title{VLT X-shooter spectroscopy of the nearest brown dwarf binary\thanks{Based on observations collected at the European Southern Observatory, Chile, under DDT programme 290.C-5200(B) (PI Lodieu)}}


   \author{N. Lodieu \inst{1,2},
          M.\ R.\ Zapatero Osorio \inst{3},
          R.\ Rebolo \inst{1,2,4},
          V.\ J.\ S.\ B\'ejar \inst{1,2},
          Y.\ Pavlenko \inst{5,6},
          A.\ P\'erez-Garrido \inst{7}
          }

   \institute{Instituto de Astrof\'isica de Canarias (IAC), Calle V\'ia L\'actea s/n, E-38200 La Laguna, Tenerife, Spain. 
         \email{nlodieu,vbejar,rrl@iac.es}
         \and
         Departamento de Astrof\'isica, Universidad de La Laguna (ULL), E-38205 La Laguna, Tenerife, Spain.
         \and
         Centro de Astrobiolog\'ia (CSIC-INTA), Ctra. Ajalvir km 4, E-28850 Torrej\'on de Ardoz, Madrid, Spain.
         \email{mosorio@cab.inta-csic.es}
         \and
         Consejo Superior de Investigaciones Cient\'ificas, CSIC, Spain.
         \and
         Main Astronomical Observatory of the National Academy of Sciences of Ukraine.
         \and
         Center for Astrophysics Research, University of Hertfordshire, College Lane, Hatfield, Hertfordshire AL10 9AB, UK 
         \and
         Universidad Polit\'ecnica de Cartagena, Campus Muralla del Mar, Cartagena, E-30202 Murcia, Spain.
             }

   \date{Received \today{}; accepted \today{}}
 
  \abstract
   {}
   {The aim of the project is to characterise both components of the 
nearest brown dwarf sytem to the Sun, WISE J104915.57$-$531906.1 (also proposed
as Luhman\,16AB) at optical and near-infrared wavelengths.}
   {We obtained high signal-to-noise intermediate-resolution (R\,$\sim$\,6000--11000) 
optical (600--1000 nm) and near-infrared (1000--2480nm) spectra of each component of
Luhman\,16AB, the closest brown dwarf binary to the Sun, with the X-Shooter instrument
on the Very Large Telescope (VLT).}
   {
We classify the primary and secondary of the Luhman\,16 system as L6--L7.5 and
T0$\pm$1, respectively, in agreement with previous measurements published
in the literature. We present measurements of the lithium pseudo-equivalent
widths, which appears of similar strength on both components 
(8.2$\pm$1.0\AA{} and 8.4$\pm$1.5\AA{} for the L and T components, respectively). 
The presence of lithium ($^7$Li) in both components imply masses below 
0.06 M$_{\odot}$ while comparison with models suggests lower limits of 0.04 M$_{\odot}$.
The detection of lithium in the T component is the first of its kind.
Similarly, we assess the strength of other alkali lines (e.g.\ pseudo-equivalent widths
of 6--7\AA{} for Rb{\small{I}} and 4--7\AA{} for Cs{\small{I}}) present in the optical
and near-infrared regions and compare with estimates for L and T dwarfs. 
We also derive effective temperatures and luminosities of each component 
of the binary: $-$4.66$\pm$0.08 dex and 1305$^{+180}_{-135}$ for the L dwarf
and $-$4.68$\pm$0.13 dex and 1320$^{+185}_{-135}$ for the T dwarf, respectively. 
Using our radial velocity determinations, the binary does not appear to belong 
to any of the well-known moving group.
Our preliminary theoretical analysis of the optical and $J$-band spectra indicates 
that the L- and T-type spectra can be reproduced with a single temperature and gravity 
but different relative chemical abundances which impact strongly the spectral energy 
distribution of L/T transition objects.
}
   {}


  \authorrunning{Lodieu et al$.$}
  \titlerunning{X-shooter observations of the nearest brown dwarf binary}

   \maketitle
%

%
%
\section{Introduction}
\label{Luhman16_XSH:intro}

Since the discovery of the first brown dwarfs in 1995 \citep{nakajima95,rebolo95},
the field of substellar research has made enormous progress with the discovery
of more than 1000 nearby ultracool dwarfs, defined as objects with spectral
types later than M7, which include L 
\citep[$\sim$1300--2200\,K;][]{martin99a,kirkpatrick00,basri00,leggett00}
and T dwarfs \citep[$\sim$1300--600\,K;][]{burgasser06a}.
The coolest brown dwarfs ever found to date, originally nicknamed Y dwarfs 
\citep{kirkpatrick99} have been announced by the Wide Infrared Survey Explorer
\citep[WISE;][]{wright10} team \citep{cushing11,kirkpatrick12,tinney12}.
They have temperatures estimated to 500--300\,K and masses below 0.01
M$_{\odot}$, according to state-of-the-art models \citep{cushing11}.

%
%
\begin{table*}
 \centering
 \caption[]{Observing logs of the VLT X-shooter observations and information
on the wavelength range, resolution, and properties of the three arms of the
X-shooter spectrograph.}
 \begin{tabular}{l c c c c c c c c}
 \hline
 \hline
Arm  & Detector  & Slit  & Resolution & $\lambda$ range & ExpT & Date & Time   &  Airmas \cr
     &           & arcsec &           &  $\mu$m         &  sec & DDMMYY & hh:mm &        \cr
 \hline
UVB  & 4096$\times$2048 E2V CCD44-82    & 0\farcs8 &  6200 & 0.30--0.56 & 360 & 11062013 & 00:31--01:53 & 1.29--1.54 \cr
VIS  & 4096$\times$2048 MIT/LL CCID\,20 & 0\farcs7 & 11000 & 0.56--1.02 & 330 & 11062013 & 00:31--01:53 & 1.29--1.54 \cr
NIR  & 2096$\times$2096 Hawaii 2RG      & 0\farcs6 &  6200 & 1.02--2.48 & 6$\times$52 & 11062013 & 00:31--01:53 & 1.29--1.54 \cr
 \hline
 \label{tab_Luhman16_XSH:observing_logs}
 \end{tabular}
\end{table*}

During the past decade, $\epsilon$ Indi B was the closest brown dwarf binary 
to the Sun, located at 3.626$\pm$0.009 pc from the Sun \citep{scholz03,mjm04}.
It has a mean projected physical separation of 2.65 au (0.75 arcsec) and 
it is located at $\sim$1500 au from $\epsilon$ Indi A \citep{torres06,vanLeeuwen07}. 
It is the best studied pair of brown dwarfs with the highest quality 
dataset to date \citep{king10b}. In March 2013, \citet{luhman13a}
announced the discovery of a nearby brown dwarf with an optical
spectral of L8 at a distance of 2.00$\pm$0.15 pc, WISE J104915.57$-$531906.1A,
resolved as a close binary (physical separation of 3 au) in the $i$-band.
\citet{mamajek13a} proposed to call this new nearby brown dwarf binary Luhman\,16AB
due to its proximity to the Sun. The detection of the lithium in absorption 
at 6708\AA{} in the optical spectrum of the primary \citet{luhman13a} and
later in both components unambiguously places the system in the substellar 
regime \citep{faherty14a}. It is brighter than $\epsilon$\,Indi\,B by 
1.5 mag in $J$ and 2 mag in 
$I$. It is the third closest system to the Sun, after the Centauri system and
Barnard's star. Thus, it represents the best substellar system amenable 
for detailed characterisation of its spectral energy distribution and a 
unique target to understand the chemical processes at play at low temperatures
and accross the L/T transition \citep{burgasser14a}. This system also provides 
a rare opportunity to obtain high-resolution and high signal-to-noise spectroscopy 
at optical and infrared wavelengths to test the mass-luminosity-age relation 
predicted by state-of-the-art models as well as the chemistry, vertical mixing, 
abundances of the various species present in substellar atmospheres, the
role of clouds \citep{crossfield14}, variability \citep{gillon13a}, and 
search for planets \citep{boffin14a}.

In this paper we present medium-resolution optical and near-infrared spectra
for each component of the nearest brown dwarf binary to the Sun, Luhman\,16AB
\citep[originally known as WISE J104915.57$-$531906.1;][]{luhman13a}.
\citet{kniazev13} presented resolved optical spectroscopy covering the
670--900 nm range, while \citet{burgasser13b} published low-resolution
(120--300) near-infrared (700--2500 nm) spectra for each component. 
\citet{faherty14a} presented medium-resolution optical (R$\sim$4000) and
near-infrared (R$\sim$8000) of each component of the system during the writing
of this manuscript.
In Section \ref{Luhman16_XSH:obs}, we describe the observations conducted
in service mode with X-shooter on the European Southern Observatory (ESO) 
Very Large Telescope (VLT) and the associated data reduction. 
In Section \ref{Luhman16_XSH:SpecProp}, we assign spectral types to each component
of the system and discuss the strength of the various alkali lines, including lithium.
In Section \ref{Luhman16_XSH:Teff_Lum}, we derive physical parameters such as 
effective and luminosity for each component of the binary.
In Section \ref{Luhman16_XSH:RV}, we derive the radial velocity of the
system and discuss its membership to nearby moving groups. Synthetical spectra of the alkalis are computed and compared to the Xshooter data in Section~\ref{Luhman16_XSH:models}.

%
%
\section{Optical to infrared spectroscopy}
\label{Luhman16_XSH:obs}

We carried out spectroscopy from the UV- to the $K$-band with the X-shooter
cross-dispersed echelle spectrograph \citep{dOdorico06,vernet11} mounted 
on the Cassegrain focus of the Very Large Telescope (VLT) Unit 2\@. 
Observations were carried out in service mode on 11 June 2013 under clear 
skies and a seeing between 0.6 and 0.85 arcsec to ensure that both components 
of the binary system were resolved over the full wavelength range covered by X-Shooter.
More details on the properties of the X-shooter spectrograph and the logs
of the observations are provided in Table \ref{tab_Luhman16_XSH:observing_logs}.

The acquisition of the target was made through an optical CCD with the
Sloan $z$-band filter, which is
not problematic considering the brightness of the source. We set the
individual on-source integration times to one expossure of 360 sec, 
one of 330 sec, and six of 52 sec in the UVB, VIS, and NIR arms,
respectively, repeated five times with a maximum nod-throw of 5\arcsec~to correct 
for the sky contribution (mainly) in the near-infrared.
The field-of-view
and the slit were both rotated by a position angle of $-$47 degrees to place
both components in the slit. Hence, the observations were not conducted with 
the slit orientated along the parallactic angle. We caution the use the optical 
spectrum of each component of the binary for accurate spectral classification 
because of the effect of differential refraction particularly at high air masses.
However, the relative fluxes between components remain unaffected because both sources 
are separated by about 1\farcs5 and have similar spectral types, hence, they are 
supposedly affected the same way.

\begin{figure*}
  \centering
  \includegraphics[width=\linewidth, angle=0]{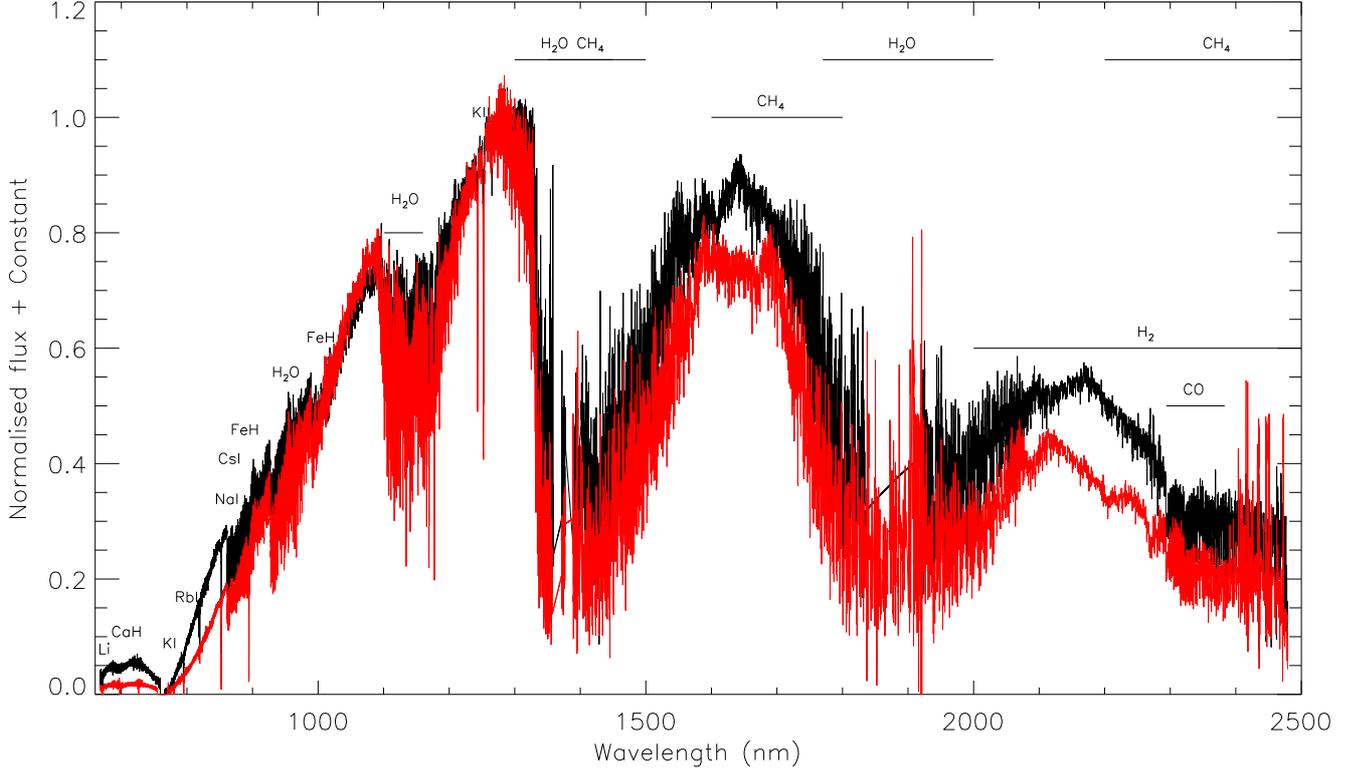}
   \caption{Comparison of the X-shooter optical and near-infrared spectra
of the L (black line) and T (red line) components of the Luhman\,16AB brown 
dwarf binary. Both spectra are normalised to unity at 1300 nm to highlight
the differences in spectroscopic features between both components.
Both spectra are also corrected for telluric contribution.
The main spectral features present in L and T dwarfs are highlighted.
   }
   \label{fig_Luhman16_XSH:spec_compare}
\end{figure*}

\begin{figure*}
  \centering
  \includegraphics[width=0.48\linewidth, angle=0]{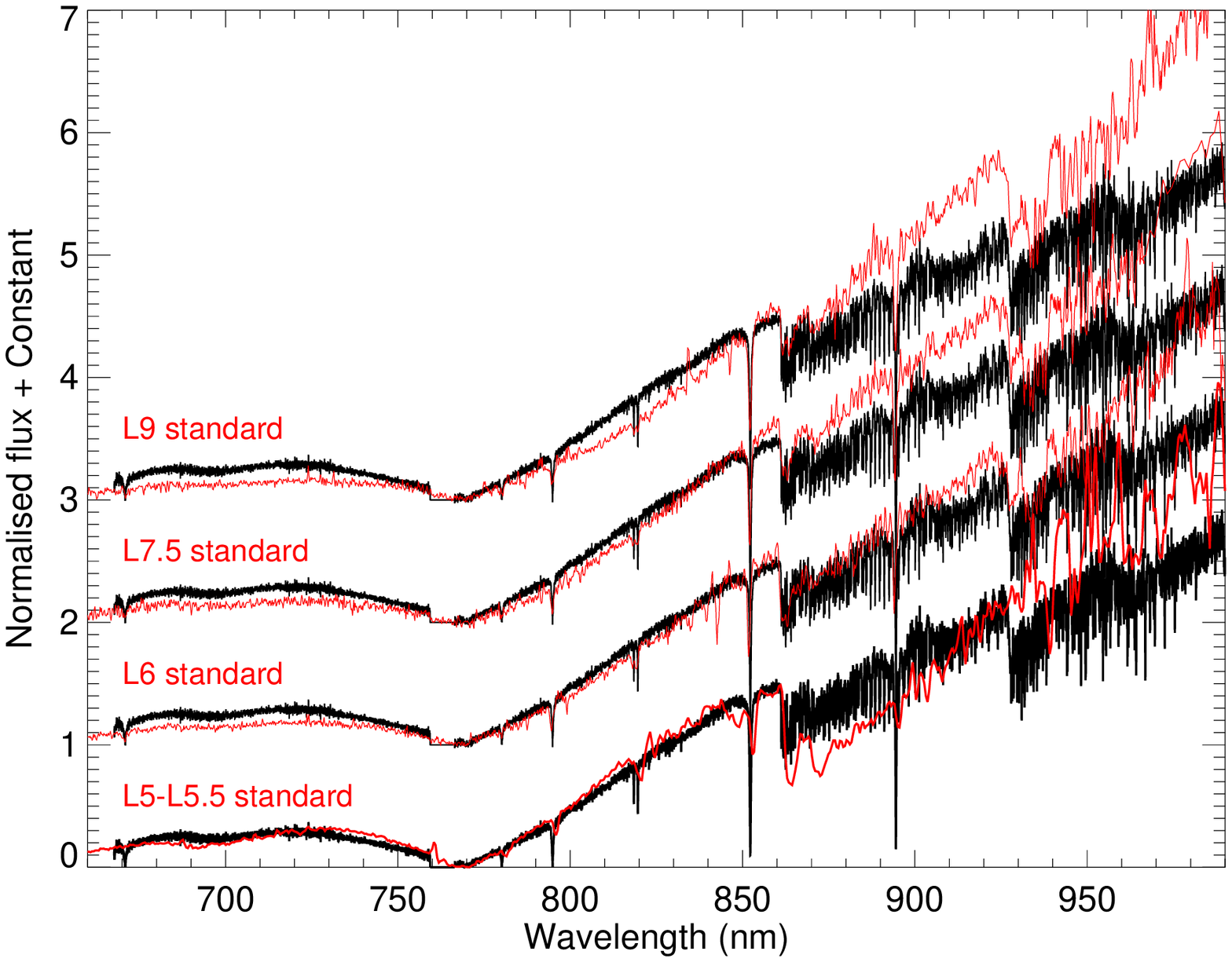}
  \includegraphics[width=0.48\linewidth, angle=0]{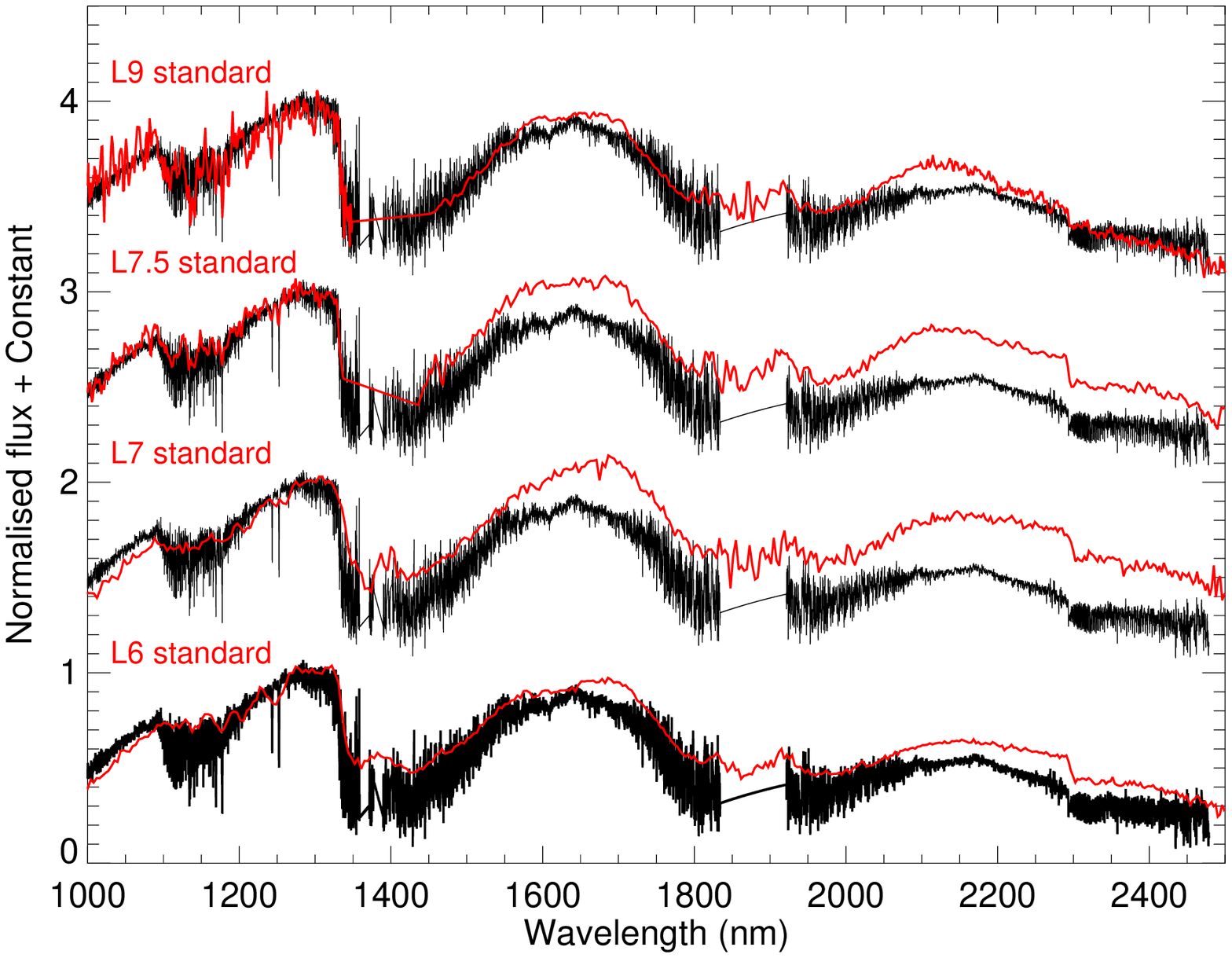}
  \includegraphics[width=0.48\linewidth, angle=0]{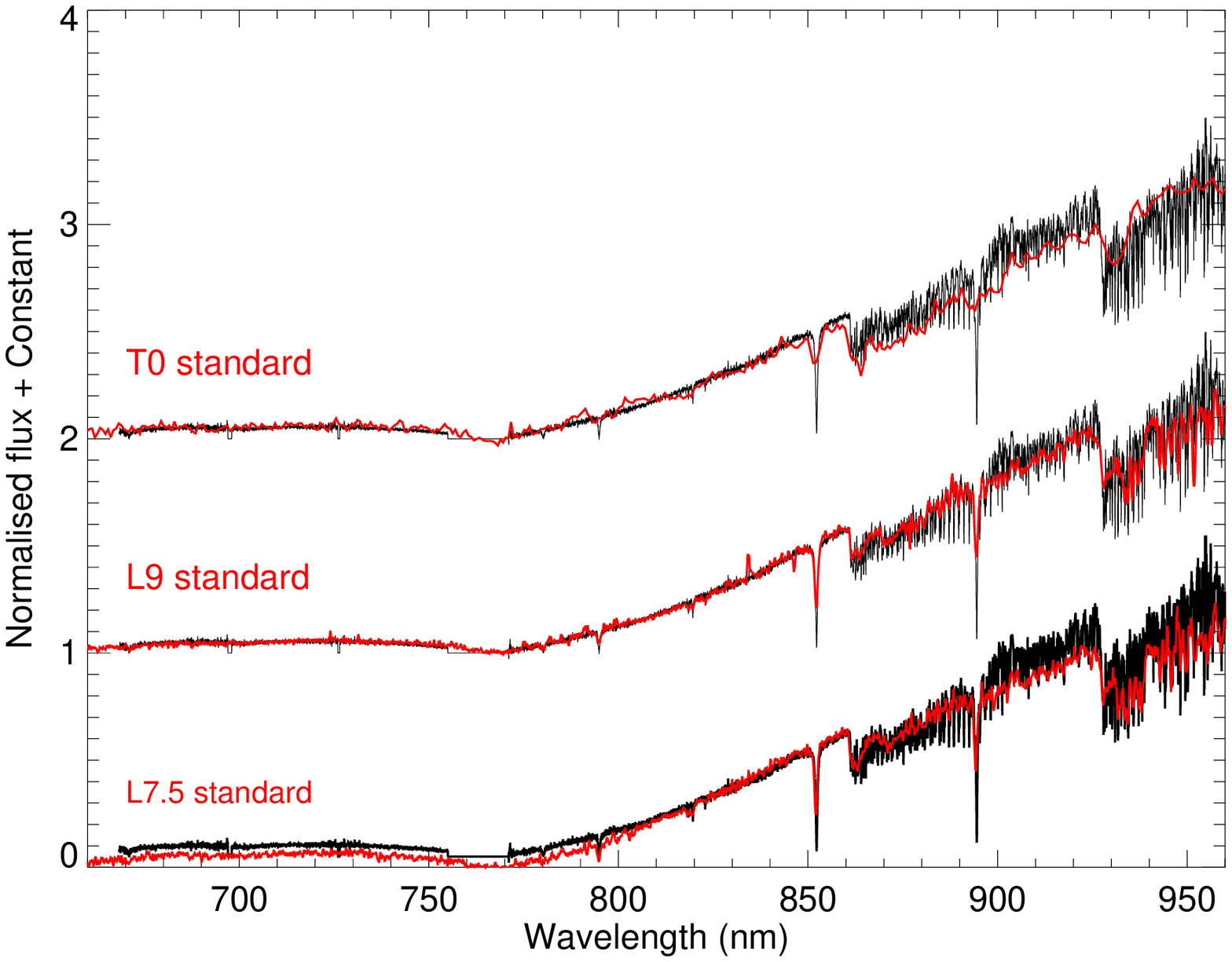}
  \includegraphics[width=0.48\linewidth, angle=0]{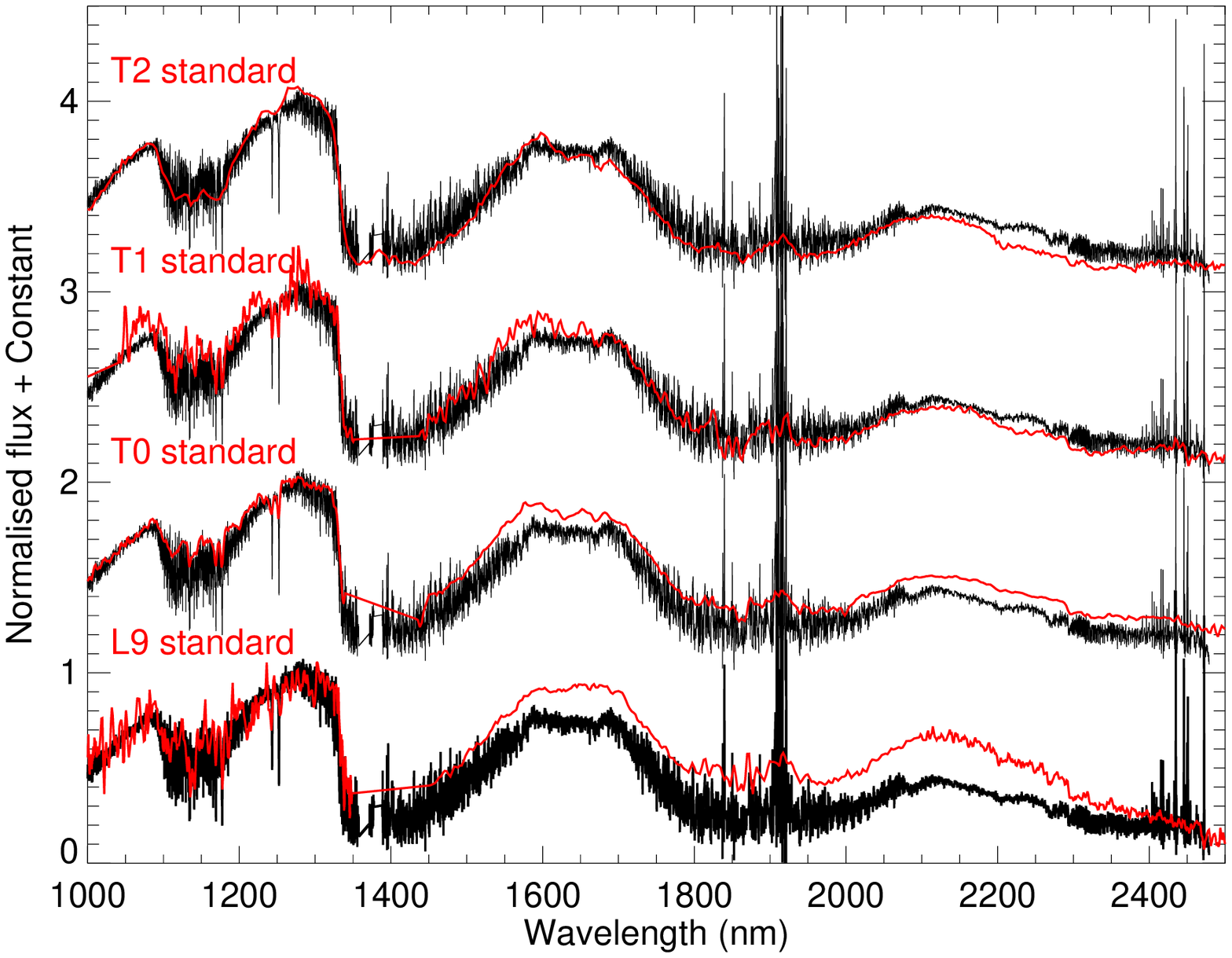}
   \caption{{\it{Top panels:}} VLT X-shooter optical (left) and near-infrared 
(right) spectra corrected for telluric lines for the L dwarf component of Luhman\,16AB\@.
{\it{Bottom panels:}} Same but for the T component of Luhman\,16AB\@.
Overplotted as red lines are spectral standards downloaded from the SpeX archive 
and Sandy Leggett's webpage:
2MASS\,J15074769$-$162738.6\citep[L5--L5.5;][]{reid00,dahn02,kirkpatrick00,knapp04},
2MASSs\,J0850359$+$105716 \citep[L6;][]{kirkpatrick99}, 
2MASSI\,J0103320$+$193536 \citep[L7;][]{kirkpatrick00},
2MASS\,J16322911$+$1904407 \citep[L7.5;][]{leggett02,geballe02},
2MASS\,J03105986$+$1648155 \citep[L9;][]{kirkpatrick00,burgasser06a,reid01a,leggett02,geballe02},
SDSS\,J04234857$-$0414035 \citep[T0;][]{leggett00,geballe02,cruz03,burgasser06a}, 
SDSS\,J08371721$-$0000180 \citep[T1;][]{leggett00,leggett02,burgasser06a,kirkpatrick08},
and SDSSp J125453.90-012247.4 \citep[T2;][]{leggett00,burgasser04c}.
All optical and near-infrared spectra are normalised at $\sim$840--860 nm
and 1300 nm to ease the comparison.
}
   \label{fig_Luhman16_XSH:spec_dL_dT}
\end{figure*}

Data reduction was performed for the VIS and NIR arms separately. For the present paper, we discarded the UVB arm because no obvious signal is registered from any of the components. We applied standard procedures for optical and visible wavelengths, which include sky subtraction and flat-field correction. All steps were carried out with routines within the IRAF\footnote{IRAF is distributed by National Optical Astronomy Observatories, whcih is operated by the Association of Universities for Research in Astronomy, Inc., under contract with the National Science Foundation.} (Image Reduction and Analysis Facility) environment \citep{tody86,tody93}. Spectra were optimally extracted with the APALL command of the ECHELLE package in IRAF. Each pair member was extracted separately using the same aperture size to define the extraction and background regions on the detectors. A third-order polynomial was employed per echelle order to trace the spectra across the detectors. Wavelength calibration was performed to a typical accuracy of $\pm$0.05~\AA~using ThAr (VIS) and ArKrXeNe (NIR) arcs taken with the same instrumental configuration as our targets. The spectra of the B9V star HIP\,56470 observed immediately after Luhman\,16AB with an air mass within $\pm$0.1 were used for division into the corresponding VIS and NIR science spectra. We previously removed the hydrogen lines intrinsic to this hot star. Finally, we multiplied the science spectra by a blackbody of temperature 10700 K corresponding to B9V. Individual echelle orders were merged by matching the counts in common wavelength ranges from one order to the next. 
We observe a peak and a dip around 2.08--2.09 microns in the spectra
of the components of the system not seen in other T dwarfs, as a potential artefact
in our X-shooter data. Though we could not remove it, it does not affect our scientific interpretation.
as a potential artefacts. However, it does not affect our interpretation.
The other regions of the spectra is not affected by any other strong artefact(s).
The final optical and near-infrared one-dimensional spectra  
are displayed in Figures~\ref{fig_Luhman16_XSH:spec_compare} and~\ref{fig_Luhman16_XSH:spec_dL_dT}. The latter also includes known L and T dwarf templates from the literature 
\citep{knapp04,golimowski04b,burgasser04c,chiu06,looper07}.

%
%
%

%
%
%

%
%
\section{Spectroscopic properties of each component}
\label{Luhman16_XSH:SpecProp}
\subsection{Spectral types}
\label{Luhman16_XSH:SpecProp_SpT}

In Fig.\ \ref{fig_Luhman16_XSH:spec_compare}, we compare the full
spectra (660--2500 nm), normalised at $\sim$1300nm,
of the L and T components (Luhman\,16A and Luhman\,16B, respectively).
We see that the L component is relatively brighter than the
T component bluewards of 1000 nm. The overall fluxes
are very similar in the $J$-band while the L component dominates
in the $H$ and $K$ bands. We confirm that Luhman\,16A
is of earlier type than Luhman\,16B\@.

We assign spectral types to each component of the binary,
using direct comparison with known L and T spectral templates both
at optical and near-infrared wavelengths available at the
SpeX archive\footnote{http://pono.ucsd.edu/$\sim$adam/browndwarfs/spexprism/}
and Sandy Leggett's homepage\footnote{http://staff.gemini.edu/~sleggett/LTdata.html}
\citep{chiu06,golimowski04b,knapp04}. We classified each component separately
in the optical and the near-infrared.
The optical spectrum of the T component is best fit by the L9 template
2MASS\,J03105986$+$1648155 \citep{kirkpatrick00,burgasser06a,reid01a,leggett02,geballe02}
whereas its near-infrared is best reproduced by the T0 template SDSS\,J04234857$-$0414035
\citep[T0;][]{leggett00,geballe02,cruz03,burgasser06a} with an uncertainty of one subtype
(Fig.\ \ref{fig_Luhman16_XSH:spec_dL_dT}).
The L component is not well reproduced by any of the L dwarf spectral
templates over the full spectral energy range. 

Our final spectral classifications, T0$\pm$1 for Luhman\,16B and L6--L7.5 to Luhman\,16A
(Fig.\ \ref{fig_Luhman16_XSH:spec_dL_dT}), agree
with previous studies, such as \citet[][L7.5$\pm$0.5 and T0.5$\pm$0.5]{burgasser13b}
and \citet[][L8$\pm$1 and T1$\pm$2]{kniazev13}.

\subsection{Molecules}
\label{Luhman16_XSH:SpecProp_Mol}

In the optical spectra of both components, we observe a wide pressure-broadened potassium
doublet around 760 nm. We can also see the FeH, CrH, and H$_{2}$O absorption bands at
$\sim$860nm, $\sim$870 nm, and $\sim$930 nm, respectively, pointing towards cool temperatures 
and late spectral types \citep{kirkpatrick99,kirkpatrick00,reid00}.

In the near-infrared, we observe a stronger methane absorption band at around 1.1--1.2 microns
in the T component, indicating a potentially cooler temperature. By comparison, the shape of 
the spectra in the $H$-band appear very similar, pointing towards a small difference in
spectral type. We detect the onset of methane (CH$_{4}$) absorption at 2.12 $\mu$m in 
Luhman\,16B, which clearly indicates a T spectral type. Additionally, we find CO in
absorption at 2.3 $\mu$m in both L and T components. The co-existence of CH$_{4}$ and 
CO hints at a peculiar carbon chemistry.

%
%
\begin{table}
\small
\tabcolsep=0.1cm
\centering
\caption{Pseudo-equivalent widths (in \AA) of atomic features. \label{tab_Luhman16_XSH:table_EWs}}
\begin{tabular}{lccc}
\hline  \hline
\multicolumn{1}{c}{Line\tablefootmark{a}} & 
\multicolumn{1}{c}{Luhman 16A} & 
\multicolumn{1}{c}{Luhman 16B} & 
\multicolumn{1}{c}{$\Delta \lambda$\tablefootmark{b}} \\ 
\multicolumn{1}{c}{(nm)} & 
\multicolumn{1}{c}{} & 
\multicolumn{1}{c}{} & 
\multicolumn{1}{c}{(nm)} \\ 
\hline
Li\,{\sc i} $\lambda$670.782 & 8.2$\pm$1.0 & 8.4$\pm$1.5 & 669.45--673.50 \\
Rb\,{\sc i} $\lambda$780.023 & 6.2$\pm$0.7 & 6.7$\pm$0.8 & 779.04--781.44 \\
Rb\,{\sc i} $\lambda$794.760 & 6.7$\pm$0.7 & 7.1$\pm$0.7 & 793.12--796.71 \\
Na\,{\sc i} $\lambda$818.326 & 1.0$\pm$0.2 & 0.5$\pm$0.1 & 818.00--819.00 \\
Na\,{\sc i} $\lambda$819.482 & 1.3$\pm$0.2 & 0.8$\pm$0.1 & 819.00--820.15 \\
Cs\,{\sc i} $\lambda$852.115 & 6.5$\pm$0.7 & 7.4$\pm$0.7 & 849.85--854.44 \\
Cs\,{\sc i} $\lambda$894.348\tablefootmark{c} & 4.2$\pm$0.5 & 5.0$\pm$0.5 & 893.65--895.07 \\
K\,{\sc i} $\lambda$1169.022\tablefootmark{d} & 5.3$\pm$0.5 & 7.5$\pm$0.5 & 1167.93-1170.59 \\
K\,{\sc i} $\lambda$1173.284\tablefootmark{d} & 5.4$\pm$0.5 & 7.1$\pm$0.5 & 1176.34--1178.91 \\
K\,{\sc i} $\lambda$1243.227 & 3.0$\pm$0.5 & 4.6$\pm$0.5 & 1241.15--1245.00 \\
K\,{\sc i} $\lambda$1252.213 & 4.0$\pm$0.5 & 6.1$\pm$0.5 & 1250.25--1254.25 \\
\hline
\end{tabular}
\tablefoot{ 
\tablefoottext{a}{Catalogue wavelengths are given in the air system.} 
\tablefoottext{b}{Wavelength interval over which the line profile is integrated.}
\tablefoottext{c}{Blended with FeH absorption.}
\tablefoottext{d}{Blended with H$_2$O absorption.}
}
\end{table}

\subsection{Lithium 7 ($^{7}$Li)}
\label{Luhman16_XSH:SpecProp_Li7}

The presence of lithium in the atmospheres of fully convective dwarfs cooler than M6 
has been widely used as a solid criterion for substellarity 
\citep[and references therein]{rebolo92,basri00,kirkpatrick08}. 
We detect the lithium resonance doublet at 670.782 nm in both components, 
thus confirming their brown dwarf nature and masses $\le$0.060 M$_\odot$ independently of the 
age of the binary. This is the first detection of Li\,{\sc i} absorption in a T dwarf
\citep[also pointed out by][]{faherty14a}). We measured the equivalent width of the spectral 
feature (note that the doublet is not resolved) with respect to the objects' relative continuum 
or pseudo-continuum modulated by the strong absorptions due to Na\,{\sc i} to the blue and 
K\,{\sc i} to the red. We therefore refer to pseudo-equivalent widths (pEWs). The integration 
of the line profile is always performed over the range 669.45--672.25 nm. Our measurements 
and their associated uncertainties (determined as the scatter of the measurements obtained 
by placing the pseudo-continuum up and down within the photon noise of the observed spectra) 
are provided in Table\ \ref{tab_Luhman16_XSH:table_EWs}.

%
%
\begin{figure}
  \centering
  \includegraphics[width=\linewidth]{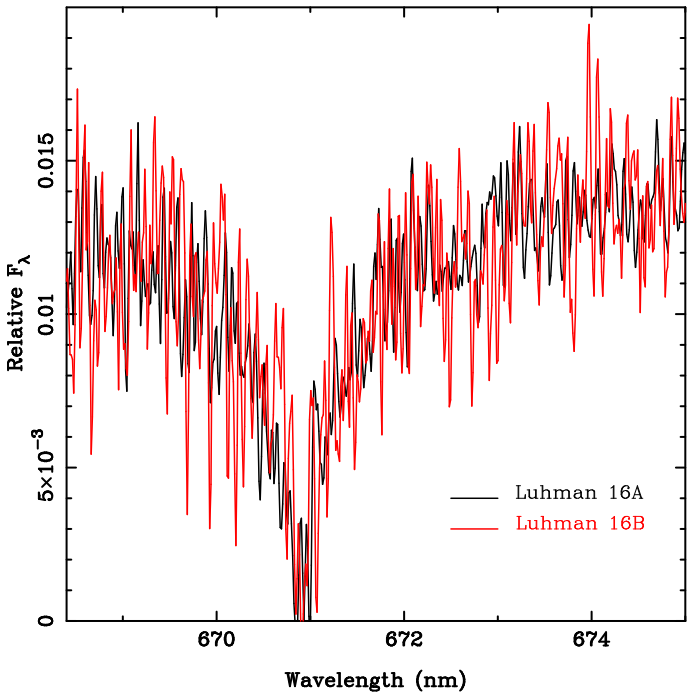}
   \caption{Spectral profiles of the Li\,{\sc i} resonance doublet at 670.782 nm seen in absorption. 
The lines of the two objects are similar in strength. The relative fluxes of both pair members 
are normalized to the mean pseudo-continuum around the Li\,{\sc i} signature for a proper 
comparison. The data are free of telluric contribution. No smoothing has been applied.}
   \label{lispec}
\end{figure}

Interestingly, we find that both the primary and the secondary have similar lithium absorption 
intensities of pEW\,$\approx$\,8--9\AA{} within the error bars despite their differing spectral 
types. Figure \ref{lispec} illustrates the Li\,{\sc i} profiles of Luhman 16A and Luhman 16B\@. 
This contrasts with the Li\,{\sc i} pEWs reported by \citet{faherty14a}. These authors used 
spectra of resolution $R$\,$\sim$\,4000 at $I$-band taken 1.5 months earlier than our X-shooter 
data and derived that the T-type component showed a lithium spectral signature smaller 
(pEW\,=\,3.8\,$\pm$\,0.4\AA{}) than that of the L-type component (pEW\,=\,8.0\,$\pm$\,0.4\AA{}). 
Whereas our pEW measurement agrees with \citet{faherty14a} result at the 1-$\sigma$ level for 
the L dwarf, there is a discrepancy by a factor of $\sim$2 for the T component. 
We repeated our procedure on the optical spectra kindly provided by Jackie Faherty and 
measured similar pEWs in Luhman\,16A and Luhman\,16B of the order of 7--8\,\AA{}, in
excellent agreement with our X-shooter pEWs. Thus, the different pEWs determinations 
are likely due to distinct measurement techniques rather than to variability. 

%
%
\begin{figure}
  \centering
  \includegraphics[width=\linewidth, angle=0]{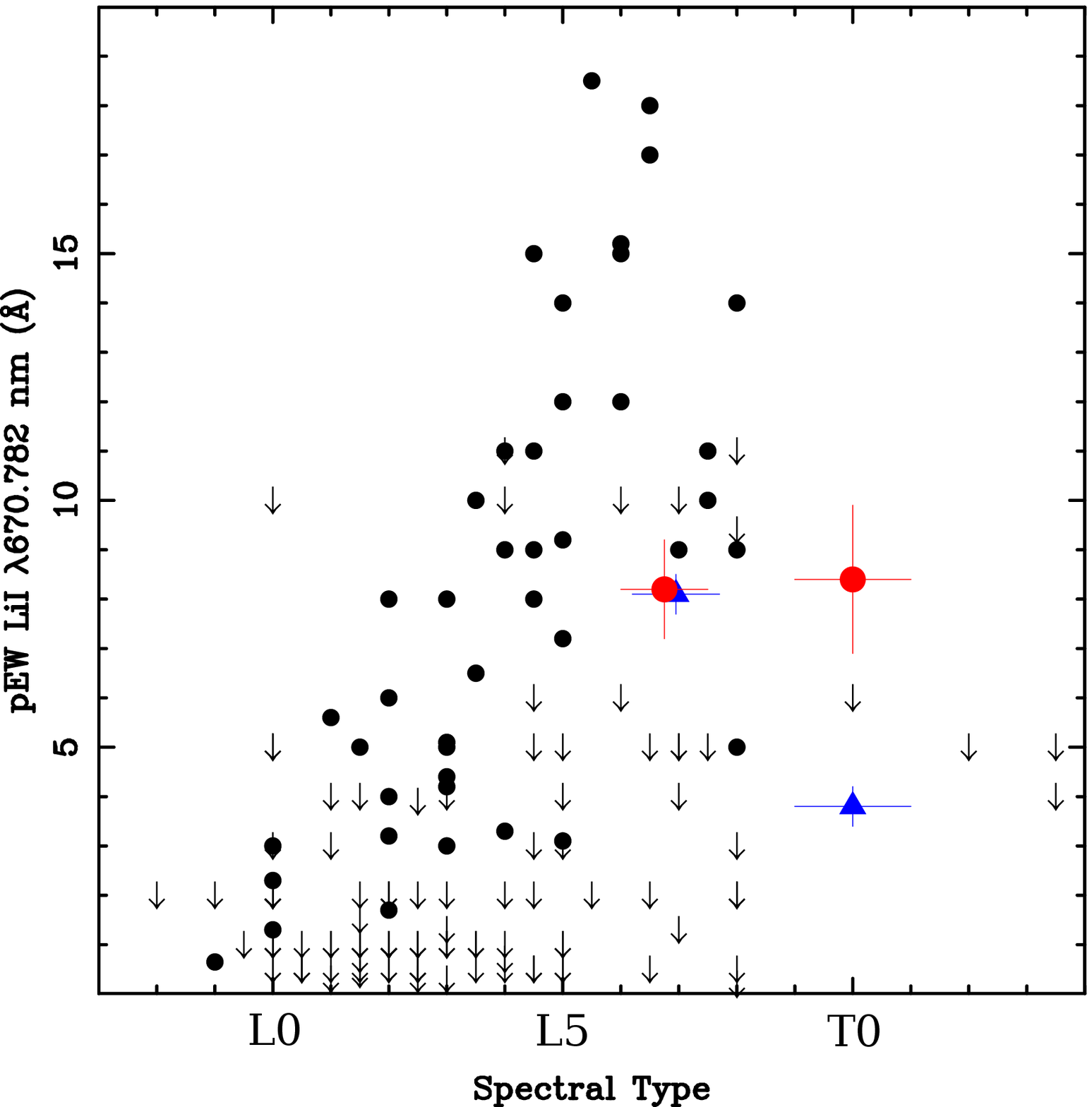}
   \caption{Li\,{\sc i} pseudo-equivalent widths (pEWs) of late-M, L, and T dwarfs. 
Data taken from \citet{tinney98a}, \citet{kirkpatrick99,kirkpatrick00,kirkpatrick08}, 
\citet{pavlenko07}, \citet{cruz09}, and \citet{zapatero14a} are shown with black dots 
(detections) and arrows (upper limits).
Our pEWs of Luhman\,16A (L6.75) and Luhman\,16B (T0) are indicated with the red dots, 
and \citet{faherty14a} measurements are plotted as blue solid triangles.
Values from \citet{faherty14a} for Luhman\,16A are slightly shifted in spectral 
type and pEW by $+$0.1 subtype and $+$0.1\AA{}, respectively, for clarity.}
   \label{fig_Luhman16_XSH:pEWs_Lithium}
\end{figure}

Our Li\,{\sc i} pEWs and those of \citet{faherty14a} are depicted as a function of spectral type in 
Fig.\ \ref{fig_Luhman16_XSH:pEWs_Lithium}. To put our values in context, we also include pEWs of 
field late-M, L, and T dwarfs given by \citet{tinney98a}, \citet{kirkpatrick99,kirkpatrick00,kirkpatrick08}, 
\citet{pavlenko07}, \citet{cruz09}, and \citet{zapatero14a}. 
As shown in Fig.\ \ref{fig_Luhman16_XSH:pEWs_Lithium}, the measurement 
of Luhman\,16A is consistent with the pEWs of L7--L8 dwarfs with lithium detections. There is a 
large scatter in Li\,{\sc i} pEWs of L dwarfs possibly due to the uncertainties associated with 
the measurements and spectral types, the unequal spectral resolutions and ways of deriving pEWs 
by the various groups, distinct states of lithium depletion, variability, and/or diverse metallicity. 
However, \citet{kirkpatrick00,kirkpatrick08} discussed that the strength of the Li\,{\sc i} 
670.782 nm line increases with later spectral type, peaks at L5--L6, and declines toward cooler 
types. As predicted by the theory of stellar and substellar chemistry (e.g., \citealt{burrows99,lodders06}), 
lithium is in atomic form above $\sim$1500--1700 K at 1 bar pressure but is converted into molecules 
(LiOH, LiCl, LiH, \ldots{}) at lower temperatures. The resulting effect on the observed spectral 
energy distribution is the depletion of the atomic lithium abundance and the decreasing intensity 
of the Li\,{\sc i} absorption at 670.782 nm toward  very low temperatures and late spectral types. 
Nevertheless, this feature is still present and remains detectable with relatively high strengths 
in early-Ts, as demonstrated by our observations of Luhman~16B and in agreement with the predictions 
made by \citet{pavlenko00a}, \citet{allard01}, and \citet{burrows02a}. Furthermore, the 
pEW\,=\,8.4\,$\pm$\,1.5\AA{} of Luhman\,16B is compatible with the computed intensity of the 
Li\,{\sc i} resonance doublet for $T_{\rm eff}$ = 1000--1200 K, full lithium preservation, and 
moderate dust opacity given in Table 5 of \citet{pavlenko00a}. 

Among the Li\,{\sc i} subordinate lines, the one at 812.64 nm would be more easily detectable because 
there is more flux (better signal-to-noise data) in our targets than at 610.36 nm. The detection of 
one of these lines would provide a direct evidence of the full preservation of lithium in the dwarfs' 
atmospheres, since the formation of the subordinate lines is extremely sensitive to the atomic abundance. 
Unfortunately, the detection of these lines in absorption also depends on the presence of dust opacity, 
which considerably reduces the line strength. We can impose an upper limit of 30 m\AA{} on the Li\,{\sc i} 
$\lambda$812.64 nm pEW for both components. A detailed spectral analysis of the binary data would 
be required to confirm whether lithium is fully preserved or depleted in both components.

%
%
\begin{figure}
  \centering
  \includegraphics[width=\linewidth, angle=0]{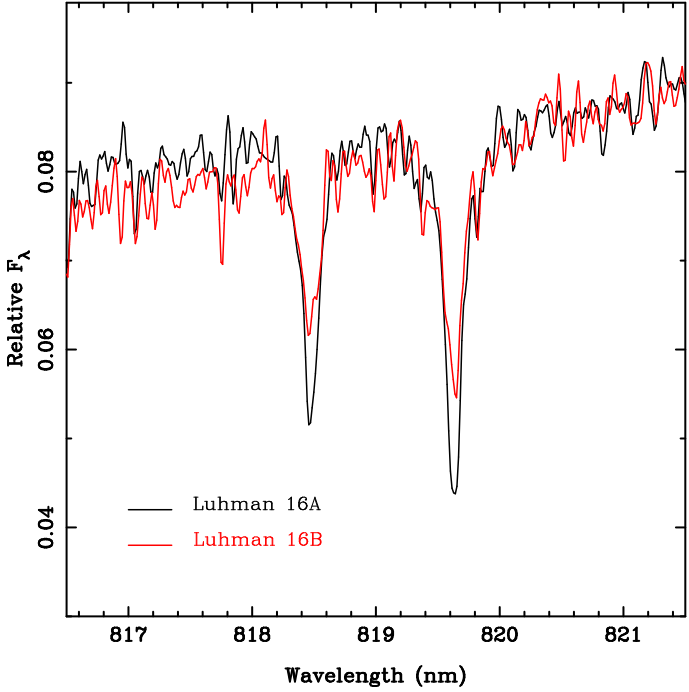}
   \caption{Spectral profiles of Na\,{\sc i} $\lambda$818.326 and $\lambda$819.482 nm 
subordinate lines. The relative fluxes of the two objects have been normalized to the 
mean pseudo-continuum redwards of $\lambda$819.482 nm. Spectra are telluric corrected 
and no smoothing is applied. }
   \label{naspec}
\end{figure}
%

%
%
\begin{figure}
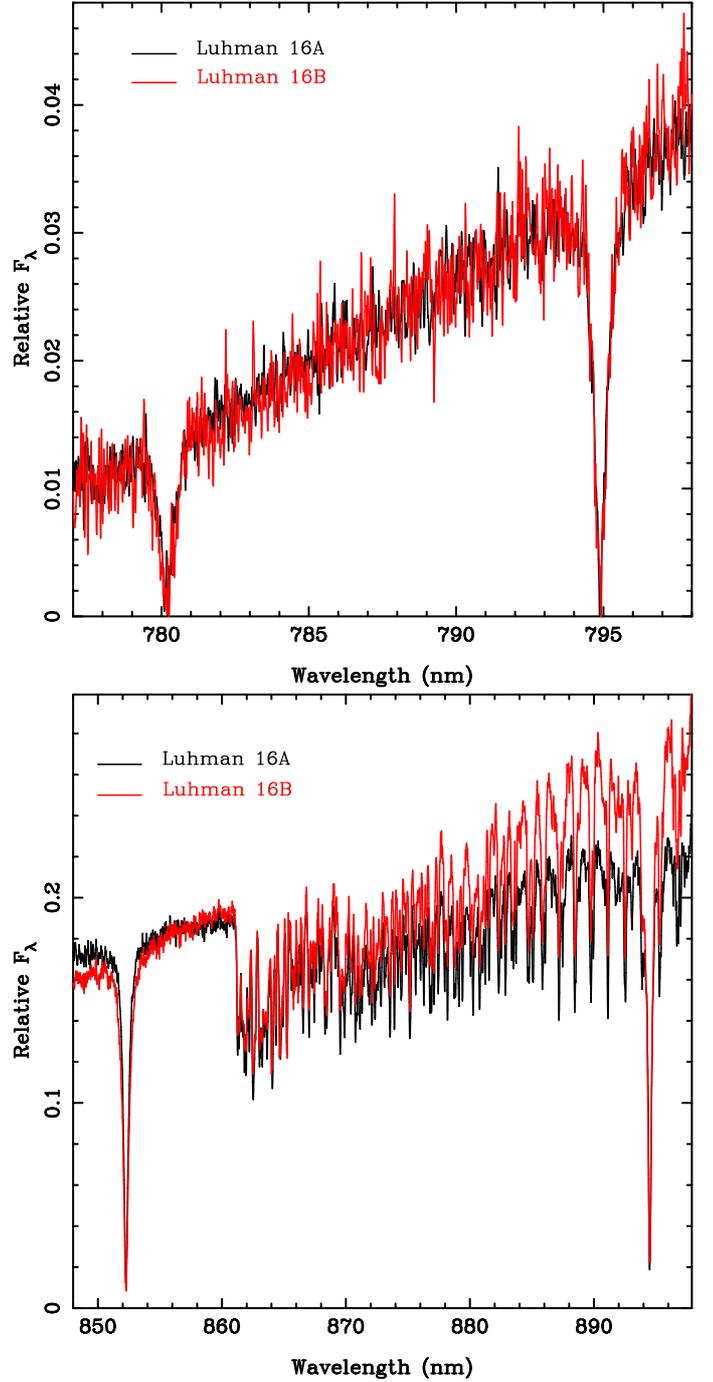

  \centering
  \includegraphics[width=\linewidth, angle=0]{24933fig9.ps}
  \includegraphics[width=\linewidth, angle=0]{24933figA.ps}
   \caption{Spectral profiles of Rb\,{\sc i} (top) and Cs\,{\sc i} (bottom) atomic lines of 
Luhman\,16AB. The relative fluxes of the two objects have been normalized to the mean 
pseudo-continuum around Rb\,{\sc i} and the Cs\,{\sc i} $\lambda$852.11 nm lines for a 
proper comparison. These data are corrected for telluric contribution. The effect of the 
different $T_{\rm eff}$ between the components becomes apparent in the slope of the 
pseudo-continua in the bottom panel. The Cs\,{\sc i} $\lambda$894.35 nm line is blended 
with the absorption due to CrH and FeH.}
   \label{rbcs}
\end{figure}
%

%
%
\begin{figure}
  \centering
  \includegraphics[width=\linewidth, angle=0]{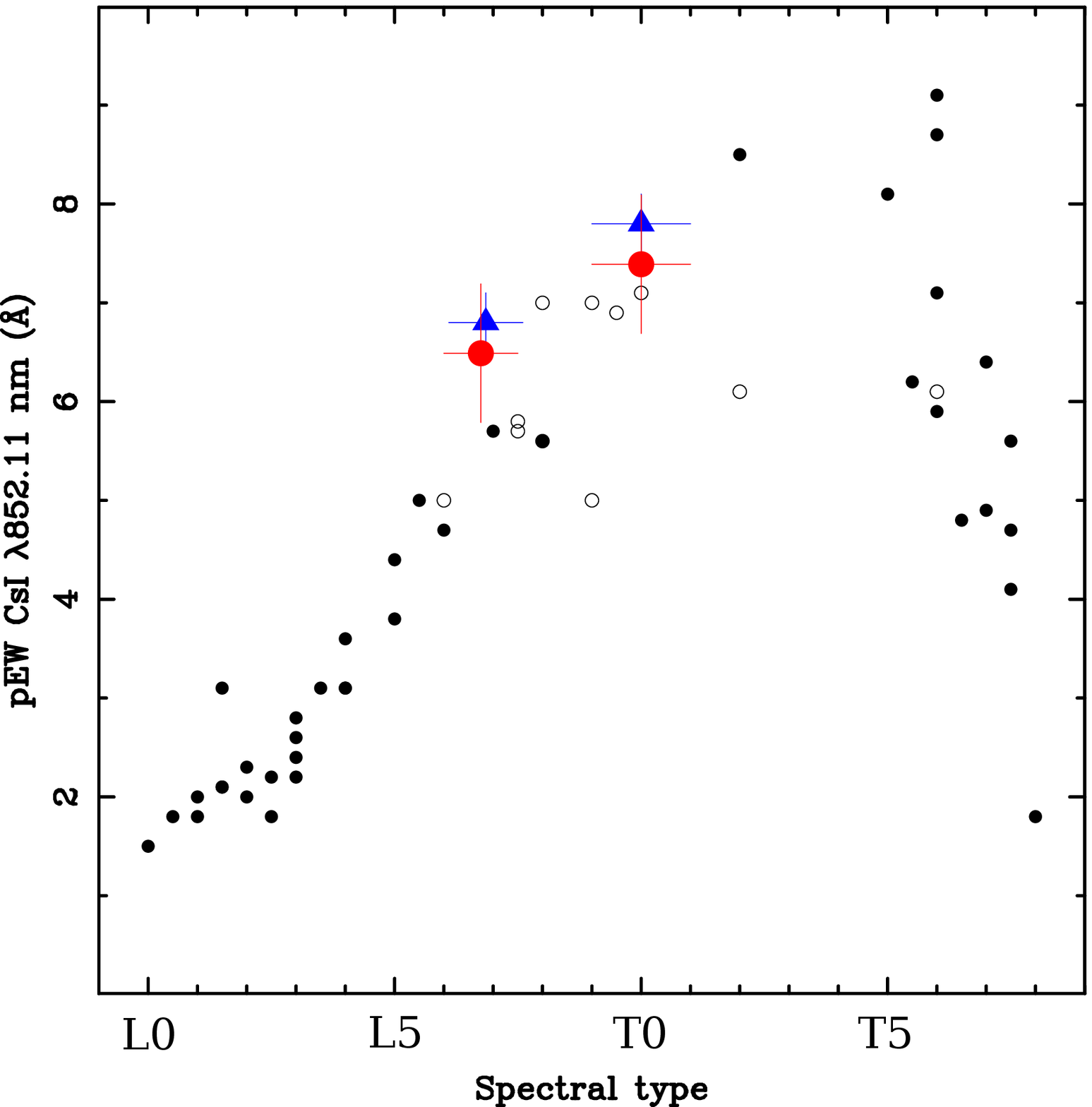}
   \caption{Cs\,{\sc i} $\lambda$852.115 nm pseudo-equivalent widths (pEWs) of L and T dwarfs. 
Our pEWs of Luhman\,16A (L7.5) and Luhman\,16B (T0) are indicated with the red dots, and \citet{faherty14a} 
measurements are plotted as blue solid triangles. Data taken from \citet{kirkpatrick00} and 
\citet{burgasser03d} are depicted with black solid circles, and our measurements over optical 
spectra available in the literature are plotted as open circles.}
   \label{cspew}
\end{figure}
\subsection{Lithium 6 ($^{6}$Li)}
\label{Luhman16_XSH:SpecProp_Li6}

Altough the high superficial gravity in brown dwarfs produces a large broadening of the alkali lines, 
in particular of the $^{7}$Li resonance doublet, it is worth to study the potential presence of the 
more fragile $^{6}$Li isotope in low-mass brown dwarfs. According to theoretical computations 
\citep[e.g.][]{nelson93b}, $^{6}$Li, which is burnt at lower temperatures than $^{7}$Li, would be 
fully preserved in brown dwarfs of mass lower than 40 M$_{\rm Jup}$. Given the potential masses 
and the brightness of the Luhman\,16AB binary, these brown dwarfs offer the interesting opportunity 
to investigate the presence of $^{6}$Li\@. The isotopic shift between the doublets of $^{7}$Li and 
$^{6}$Li is only of 0.16\AA{} and therefore requires a fine spectral resolution and high signal-to-noise.  

A detection of $^{6}$Li would clearly hint to a mass below 40 M$_{\rm Jup}$. The lithium isotopic 
ratio in meteorites is $^{7}$Li/$^{6}$Li\,$\sim$\,12, and estimates in the interstellar medium 
give similar values \citep{lemoine93}, so the presence of the $^{6}$Li isotope is expected to 
introduce a  weak asymmetry in the red wing of the lithium doublet. We investigated if there 
is any evidence for such asymmetry in the spectrum of Luhman\,16A (with a better signal-to-noise
than Luhman\,16B) by subtracting a Voigt and a Gaussian function from the observed profile.
We found flat subtractions with no significant deviation (particularly in the red wing of the 
absorption) at the level of 1/8 of the Li relative flux (1-$\sigma$) that could be attributed to 
the presence of $^{6}$Li. Higher spectral resolution and improved signal-to-noise spectroscopy
of the lithium feature is required to impose significant constraints on the presence of the 
lightest lithium isotope and in turn on the masses of these brown dwarfs.

\subsection{Other alkali lines}
\label{Luhman16_XSH:SpecProp_Alkali}

Rb\,{\sc i} at 780.023 and 794.760 nm, and Cs\,{\sc i} at 852.115 and 894.348 nm  are clearly 
seen in the X-shooter optical spectra with pEWs given in Table~\ref{tab_Luhman16_XSH:table_EWs}. 
We measured these pEWs and their error bars in the same manner as for lithium. Fig.\ \ref{rbcs} 
displays an enlargement of the X-shooter spectra around the Rb\,{\sc i} and Cs\,{\sc i} lines; 
we do not smooth the data for this comparison. The intensity of the Rb\,{\sc i} atomic signatures 
is quite similar within the error bars for both the L and T components (see top panel of 
Fig.\ \ref{rbcs}); however, Cs\,{\sc i} appears slightly stronger in the T dwarf (bottom  panel 
of Fig.\ \ref{rbcs}), which may be a signpost of its cooler temperature. 

Contrary to the Li\,{\sc i} resonance line (Section~\ref{Luhman16_XSH:SpecProp_Li7}), our pEWs of Rb\,{\sc i} 
and Cs\,{\sc i} agree with the equivalent widths published by \citet{faherty14a} within 1-$\sigma$ 
the uncertainty. The measured pEWs are also consistent with the values found for late-Ls and early-Ts 
in the literature \citep{kirkpatrick00,burgasser03d}, as it is illustrated in Fig.\ \ref{cspew}, 
where the strength of Cs\,{\sc i} $\lambda$852.115 nm is shown as a function of spectral type. 
We choose this Cs\,{\sc i} feature because it is located far red from the strong K\,{\sc i} 
absorption in the optical and it is not blended with other atomic or molecular absorptions. 
To complete the displayed data, we measured pEWs over optical spectra of field L6--T6 dwarfs 
available to us and published by \citet{kirkpatrick99,kirkpatrick00}, \citet{burgasser00b}, \citet{reid01a}, 
and \citet{cruz03}. These measurements are plotted as open circles in Fig.\ \ref{cspew}. Similarly 
to what is observed for lithium, the intensity of the cesium line rapidly increases toward later 
spectral type and reaches a maximum between T0 and T5, this is at temperatures cooler than the 
lithium case. The marked intensity decrease at $\ge$T5 is likely due to the conversion of atomic 
Cs\,{\sc i} into molecules as a natural result of the cool atmospheres' chemistry. This property 
is in good agreement with the theory, which predicts that rubidium and cesium disappears in 
monatomic form at sufficiently low temperatures that are 200--300 K smaller than the one for 
lithium (e.g., \citealt{burrows99,lodders06}). 

We detect Na\,{\sc i} $\lambda$818.326 and $\lambda$819.482 nm subordinate lines in the spectra of 
Luhman\,16A and Luhman\,16B\@. The detection is highlighted in Fig.\ \ref{naspec}. We report their pEWs in 
Table\ \ref{tab_Luhman16_XSH:table_EWs}. Both components have Na\,{\sc i} pEWs greater than 0.5\AA{}, 
which contrasts with the claim by \citet{faherty14a}. These authors imposed an upper limit on the 
strength of Na\,{\sc i} absorption of $<$0.5\AA{}. This discrepancy can be explained by the different 
spectral resolution of the data and the fact that \citet{faherty14a} spectra are not corrected for 
fringing and telluric contribution (important at these wavelengths), as acknowledged by the authors. 
Additionally, the optical spectra of \citet{kniazev13} are not free of telluric absorption; 
therefore, these authors were not able to detect this feature despite the good quality of their data. 
The L component shows stronger Na\,{\sc i} lines than the T member of the pair, probably because 
the pseudo-continuum of the latter object is more depressed due to a lower temperature and a more 
vigorous red wing of the K\,{\sc i} doublet centered on 768.2 nm. Interestingly, the presence of 
Na\,{\sc i} at 1138.145 and 1140.378 nm in the near-infrared is not obvious in either pair member, 
in agreement with \citet{faherty14a}. Other early T dwarfs, like $\epsilon$ Indi Ba, do not show 
these lines in their high quality spectra \citep{king10b}. Thus, we do not confirm the near-infrared 
detection of Na\,{\sc i} reported by \citet{burgasser13b}. If any, only the reddest line of the 
two (1140.378 nm) may display a marginally resolved signature with pEWs = 1.3\AA{} (L) and 1.5\AA{} (T), 
and no trace of the blue line. Note that this region is severely affected by H$_2$O absorption. 

Potassium is detected at optical and near-infrared wavelengths. The resonance doublet at 766.4911 
and 769.8974 nm extends several 1000\AA{} (typical of late-L and T dwarfs) and imprints the overall 
shape of the visible spectra. We report the pEWs of the near-infrared K\,{\sc i} lines in 
Table\ \ref{tab_Luhman16_XSH:table_EWs}. The strength of the optical and near-infrared 
K\,{\sc i} lines (766.4911 and 769.8974 nm, 1.169, 1.173, 1.243, and .1252 $\mu$m) is systematically 
larger in Luhman~16B than in the L dwarf, which may suggest a higher potassium abundance, a lower 
temperature, depletion of potassium in condensates, and/or a pseudo-continuum affected by 
less opacity at near-infrared wavelengths. 
The properties of the near-infrared K\,{\sc i} lines are widely discussed in comparison with 
the spectra of other field L and T dwarfs by \citet{faherty14a}. 

It is known that the alkali lines have a profound correlation with gravity (atmospheric pressure) 
and metallicity at the low temperatures of the binary: the higher the gravity and the atomic 
abundance, the stronger the spectral features. With the only exception of lithium, the similarity 
of the strength of the alkali lines between Luhman~16A and~B and field ultra-cool dwarfs of related 
spectral types indicates that the components do not deviate significantly in age and metallicity 
with respect to the field, in agreement with the age derivations by \citet{burgasser13b} and \citet{faherty14a}.

%
%
\begin{table*}
 \centering
 \caption[]{
Masses, radii, and effective temperatures (T$_{\rm eff}$) for each component
of Luhman\,16 for different ages along with the error bars from the interpolation 
of the models. \\
Note: $^{a}$ this age yields a mass that is not compatible with lithium preservation
}
 \begin{tabular}{l c c c c c c}
 \hline
 \hline
Age     & \multicolumn{3}{c}{Luhman\,16A}  & \multicolumn{3}{c}{Luhman\,16B}  \cr
        & Mass & Radius & T$_{\rm eff}$    &  Mass & Radius & T$_{\rm eff}$   \cr
Gyr     & M$_{\odot}$ & R$_{\odot}$ & K    &  M$_{\odot}$ & R$_{\odot}$ & K   \cr
 \hline
0.5       & 0.028$\pm$0.003 & 0.102$\pm$0.005 & 1223$\pm$60 & 0.029$\pm$0.003 & 0.102$\pm$0.005 & 1238$\pm$60 \cr
1         & 0.041$\pm$0.003 & 0.093$\pm$0.005 & 1280$\pm$60 & 0.041$\pm$0.003 & 0.093$\pm$0.005 & 1296$\pm$60 \cr
2         & 0.055$\pm$0.003 & 0.086$\pm$0.005 & 1328$\pm$60 & 0.055$\pm$0.003 & 0.086$\pm$0.005 & 1344$\pm$60 \cr
5\,$^{a}$ & 0.067$\pm$0.003 & 0.082$\pm$0.005 & 1368$\pm$60 & 0.067$\pm$0.003 & 0.082$\pm$0.005 & 1383$\pm$60 \cr
 \hline
 \label{tab_Luhman16_XSH:values_Mass_Radius_Teff}
 \end{tabular}
\end{table*}
%

%
%
\section{Effective temperatures and luminosities}
\label{Luhman16_XSH:Teff_Lum}
We determined the luminosity of the L and T components using $J$-, $H$-, and $K$-band bolometric corrections
\citep{golimowski04a,nakajima04a,vrba04} valid for field dwarfs and the spectral intervals L6--L8 (Luhman\,16A)
and L9--T1 (Luhman\,16B), which accounts for the spectral types of the pair members and their associated 
uncertainties. Note that the $H$-band bolometric correction (BC) shows the smallest dispersion among the 
three bands. We also used a solar bolometric luminosity of 4.73 mag and the trigonometric distance of 
2.02\,$\pm$\,0.02 pc determined for the system by \citet{boffin14a}. The near-infrared magnitudes of each 
pair member provided by \citet{burgasser13b} and \citet{kniazev13} were averaged. We derived the following 
values: log\,$L/L_\odot$ = $-4.68 \pm 0.08$ (L component) and $-4.66 \pm 0.08$ dex (T component).
The luminosity error bars consider the uncertainties in the photometry (typically $\pm$0.03 mag), the distance
modulus ($\pm$0.02 mag), and the corresponding BCs (typically $\pm$0.15 mag for $J$ and $K$, and
$\pm$0.08 for $H$). Our determination for the L component broadly agrees with \citet{faherty14a}, who found
log\,$L/L_\odot$ = $-4.67 \pm 0.04$ dex after the integration over their optical (0.6--0.9 $\mu$m) and 
near-infrared (0.8--2.5 $\mu$m) spectra supplemented with synthetic  spectra for wavelengths longer than 
2.5 $\mu$m. This supports the reliability of the BCs for these spectral types. Regarding the T component, 
our luminosity determination is consistent with that of \citet[$-4.71 \pm 0.10$ dex]{faherty14a} at the 
1-$\sigma$ level, although our measurement suggests a slightly brighter luminosity.
In a similar manner as described in \citet{faherty14a}, we also integrated our X-shooter spectra and found
that the L and T dwarfs have nearly identical luminosity, with the T object being 0.01 dex more luminous 
than the L dwarf, in agreement with the luminosities inferred from the BCs. The T component has been reported 
to show larger variability than the L source at optical and near-infrared wavelengths with peak-to-peak 
amplitudes of 11--13.5\,\%~in continuous observations and strong night-to-night evolution according to 
\citet{gillon13a}, \citet{biller13b}, and \citet{burgasser14a}. This might impact the luminosity 
determination for this particular dwarf. To account for the observed variability amplitude, an 
uncertainty of $\pm$0.05 dex should be added to the error bar in the luminosity derivation of the 
T pair member, thus yielding log\,$L/L_\odot$ = $-4.66 \pm 0.13$ dex. One striking result is that, 
despite the differing spectral classifications, both objects, Luhman~16A and~B, have consistent 
$L/L_\odot$ values at the 1-$\sigma$ level.

There are several methods to determine the effective temperatures ($T_{\rm eff}$) of Luhman~16A and~B. 
One is fitting observed spectra using theoretical model atmospheres, as it was done by \citet{faherty14a}. 

%
%
%
\begin{figure}
  \centering
  \includegraphics[width=\linewidth, angle=0]{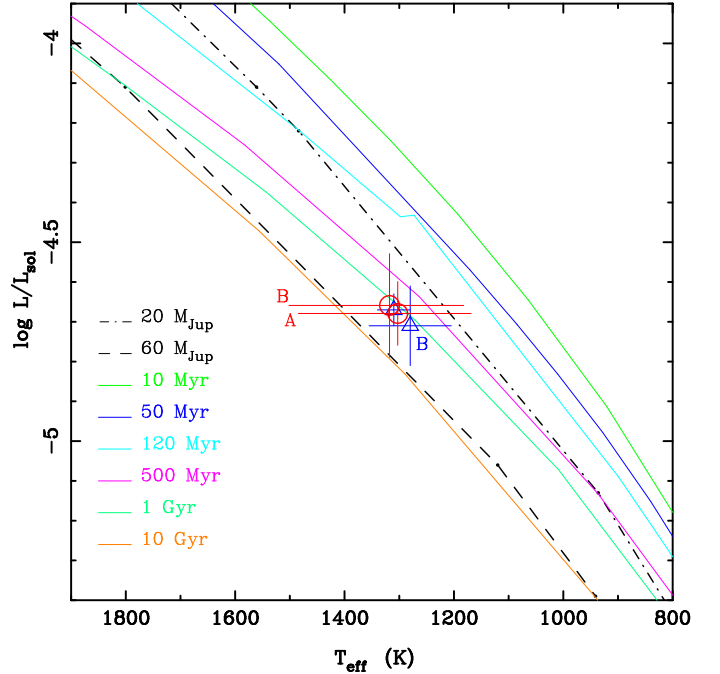}
   \caption{Luminosity versus effective temperature diagram for the two
components of the Luhman\,16AB pair (red open circles). Isochrones are plotted as solid lines, and the tracks of 0.020 and 0.060 M$_\odot$ are displayed with dot-dashed and dashed lines. The models by \citet{baraffe03} are used. Blue triangles stand for \citet{faherty14a} determinations.
}
   \label{fig_Luhman16_XSH:diagram_Teff_Lum}
\end{figure}

These authors found a wide range of possible temperatures (900--1800 K) for both components, and that 
diverse models yield temperatures that can differ by 500 K for the same source and observed data. 
Another way is to use $T_{\rm eff}$ calibrations as a function of spectral type or colour available in the 
literature \citep[e.g.][]{vrba04,stephens09}. Using these two calibrations valid for field dwarfs, we inferred 
$T_{\rm eff}$ = 1460\,$\pm$\,100 K for the L component, and 1335\,$\pm$\,100 K for the T component. 
The uncertainties stand for the dispersion of the calibrations as indicated by the authors. A third method 
consists of applying the astrophysical equation
\begin{equation}
\label{eq}
{\rm log}\,L/L_\odot = 2\,{\rm log}\,(R/R_\odot) + 4\,{\rm log}\,(T_{\rm eff}/T_{\rm eff,\odot}),
\end{equation}
where $T_{\rm eff,\odot}$ = 5777 K and $R/R_\odot$ indicates the object's radius in solar units. This way 
has the advantage that substellar objects of ages greater than $\sim$0.5 Gyr change little in size according 
to the theory of substellar evolution \citep[see review by][]{chabrier00a}. Particularly, for ages $\geq$1 Gyr 
and brown dwarfs with masses $\geq$0.02 M$_\odot$, models predict $R/R_\odot$ to be in the range 0.076--0.102\@. 
Considering a mean size of 0.09\,R$_\odot$ and the bolometric luminosities given above, we determined 
$T_{\rm eff}$ = 1305\,$^{+180}_{-135}$ for the L component and 1320\,$^{+185}_{-135}$ for the T component. 
The generous error bars include the uncertainty in luminosity and all possible radii between 0.076 and 
0.102 R$_\odot$. These error bars would decrease by about 100\,K if we would only consider the 
uncertainties in size.
The determination of the relative luminosity from the integration of the spectra has a relative error
estimated at +/-0.04 dex, which translates into a relative error on the $T_{\rm eff}$ of $\pm$2.3\% 
or $\pm$30\,K at 1300\,K\@. Hence, both objects have similar luminosity within +/-0.04 dex and similar 
$T_{\rm eff}$ within 30\,K (1$\sigma$).
We do not expect different masses or sizes from the theory if the two objects are coeval, which we 
assumed here. We list in Table \ref{tab_Luhman16_XSH:values_Mass_Radius_Teff} the masses, radii, 
and $T_{\rm eff}$ for each component of the Luhman\,16 system, assuming four different ages
(0.5, 1, 2, and 5 Gyr) along with the error bars from the interpolations of models for each age.
We note that the oldest age yields a mass that is not compatible 
with lithium preservation as already discussed by \citet{burgasser14a}.
Equation~\ref{eq} and the literature $T_{\rm eff}$---spectral type calibrations provide similar temperatures for 
Luhman\,16B, while Eq.~\ref{eq} yields a cooler temperature for Luhman\,16A, which may imply a later spectral 
classification (e.g.\ L8) as the one quoted in by \citep{luhman13a} and \citet{kniazev13}. Independently of the 
method used, we found that the two pair members have consistent temperatures and luminosities, which 
indicates similar masses and radii.

We show the bolometric luminosity versus $T_{\rm eff}$ diagram for the two components in 
Figure~\ref{fig_Luhman16_XSH:diagram_Teff_Lum}. The temperatures derived using Eq.~\ref{eq} are 
employed. We overplotted several isochrones (lines in colours) from the COND models of \citet{baraffe03}, 
ranging from 10 Myr up to 10 Gyr. We also plotted isomasses for 0.020 and 0.060 M$_\odot$. As it is seen 
from the Figure, the binary appears to be older than 120 Myr with a likely age close to 1 Gyr and a mass of 
$\sim$0.040 M$_{\odot}$ for each pair member. These determinations fully agree with the presence of lithium 
in the objects' atmospheres (see Section \ref{Luhman16_XSH:SpecProp_Li7}), and with the age and mass ranges 
determined by \citet{burgasser13b} and \citet{faherty14a}.

%
%
\section{System velocities and relative radial velocity}
\label{Luhman16_XSH:RV}

We measured the central wavelengths of the various atomic lines indicated in 
Table\ \ref{tab_Luhman16_XSH:table_EWs} to estimate the radial velocity of each component.
The values of the observed central wavelengths of  Rb\,{\sc i}, Na\,{\sc i}, Cs\,{\sc i}, and 
near-infrared K\,{\sc i}  lines are compared to their nominal air 
wavelengths\footnote{http://www.nist.gov/pml/data/asd.cfm}.
The average wavelength shift lies in the 0.10--0.15 nm range, implying a 
mean observed velocity of about 46 km\,s$^{-1}$ with a most probable range of 
36--56 km\,s$^{-1}$. The measurements for both components agree within 10\%. 
We also cross-correlated the X-shooter spectra of the L and T members against the 1400-K 
BT-Settl model spectrum computed for solar metallicity and surface gravity of 
log\,$g$ = 4.5 (cm\,s$^{-2}$) by \citet{allard12}. We used the wavelength interval 
775--1350 nm, which excludes regions of strong telluric absorption and poor signal-to-noise 
ratio and includes atomic and molecular absorption features due to Rb\,{\sc i}, Na\,{\sc i}, 
Cs\,{\sc i}, K\,{\sc i}, H$_2$O, CrH, and FeH. The resulting observed velocities are 
40.9\,$\pm$\,7.6 and 35.1\,$\pm$\,7.2 km\,s$^{-1}$ for the L and T components, respectively. 
We applied a correction of $-$15.8 km/s corresponding to the diurnal, lunar, and annual velocities 
during the observations on 11 June 2013 at UT\,=\,00$^{\rm h}$38$^{\rm m}$ to obtain the following 
heliocentric radial velocities: $v_h$ = 25.1\,$\pm$\,7.6 and 19.3\,$\pm$\,7.2 km\,s$^{-1}$ for 
Luhman 16A and Luhman16B\@. Our measurements and the heliocentric velocities 
(23.1 and 19.5 km\,s$^{-1}$) reported by \citet{kniazev13} agree within 1-$\sigma$ the uncertainty. 

Using our heliocentric radial velocities (mean value of 22.2\,$\pm$\,7.5 km\,s$^{-1}$ for the system), 
the recent determinations of parallax and proper motion by \citet{boffin14a}, and the equations 
of \citet{johnson87}, we derive the following Galactic space velocities: $U$\,=\,$-$18.3$\pm$2.0, 
$V$\,=\,$-$28.7$\pm$7.2, and $W$\,=\,$-$6.8$\pm$0.7 km\,s$^{-1}$. The uncertainties associated to 
all three Galactic velocities come from the proper motion, parallax, and radial velocity errors. 
The space velocities of Luhman\,16AB are statistically consistent with the Galaxy young disk 
kinematics according to the classification made by \cite{eggen90a} and \citet{leggett92}, 
a result that is in agreement with the age estimate made by \citet{burgasser13b} and 
\citet{faherty14a} and the value reported in the previous Section. The $UVW$ velocities do not overlap 
with any close ($\leq$100 pc),  young ($\leq$600 Myr) stellar moving group listed in the catalogues of 
\citet{zuckerman04} and \citet{torres08}, which supports the conclusions made by \citet{kniazev13}. 

We also measured the relative velocity of the T component versus the L component by means of the
cross-correlation technique. We cross-correlated the spectra with and without the telluric contribution.
The regions dominated by telluric absorption were used to establish the velocity shift  corrections
to be applied for bringing the data of the two components to the same system of reference and 
obtaining an accurate relative radial velocity.
We derived relative velocities of $-$1.06\,$\pm$\,0.50 and $-$1.42\,$\pm$\,0.70 km\,s$^{-1}$
from the optical and near-infrared parts of the X-shooter spectra, respectively. The associated 
error bars are smaller than those of the heliocentric radial velocities as expected for relative 
velocities. The measured mean relative velocity is consistent with the difference between the 
heliocentric velocities given above at the 1-$\sigma$ level. Our value is also compatible with 
that ($-$2.5\,$\pm$\,1.9 km\,s$^{-1}$) of \citet{kniazev13} obtained from data that were acquired 
90 days prior to the X-shooter spectra. The non-zero relative velocity on two different occasions 
indicates that the binary orbit is likely tilted with respect to the line of sight. Precise radial 
velocity monitoring of the system is required to search for unresolved companions (including planets) 
around any component \citep{boffin14a}, to characterize the binary orbit (with an estimated orbital 
period greater than 15 years), and to determine dynamical masses of the components, which will in turn 
become useful for comparison with substellar models.

%
%
\section{Comparison with theoretical models}
\label{Luhman16_XSH:models}

\begin{figure*}
  \centering
  \includegraphics[width=0.49\linewidth, angle=0]{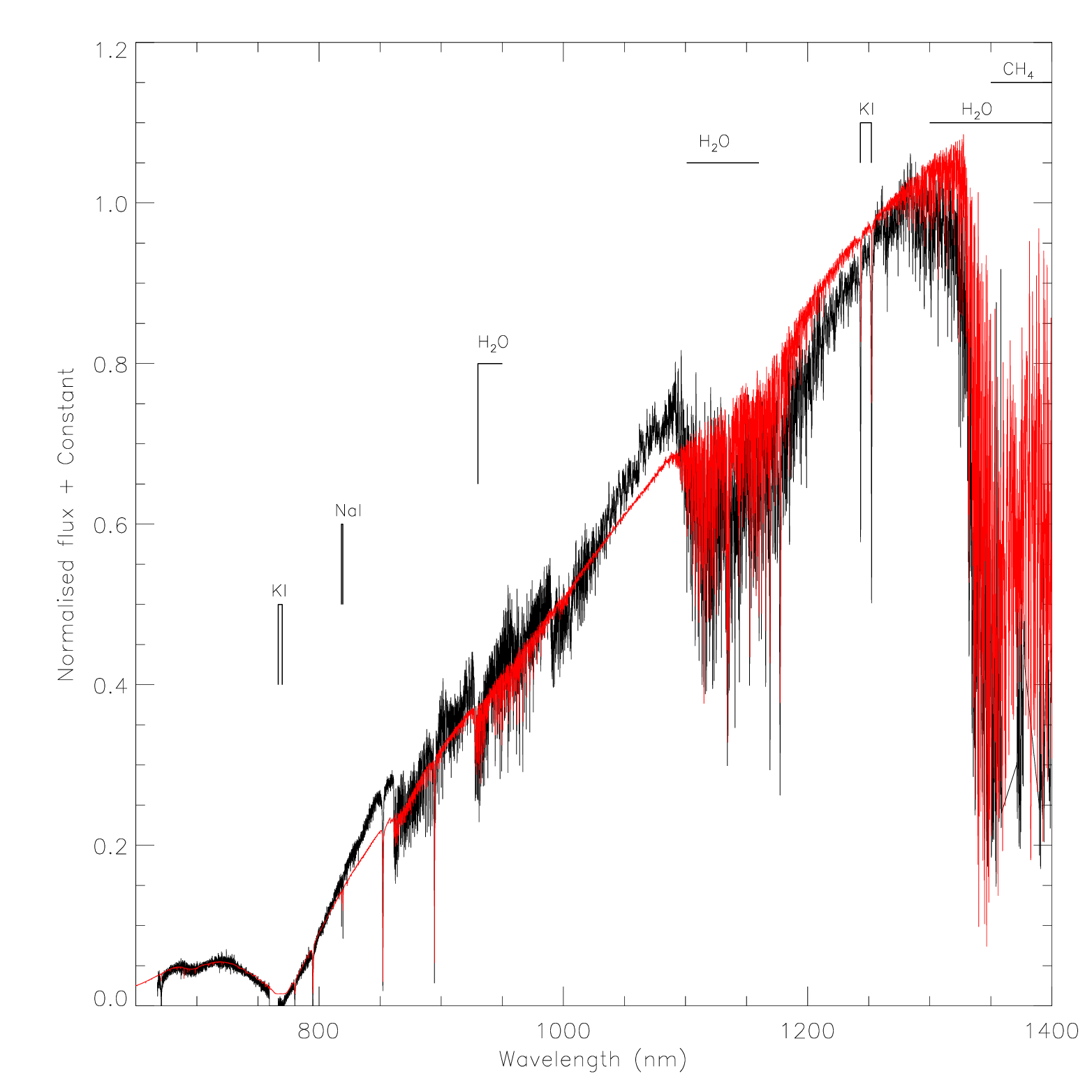}
  \includegraphics[width=0.49\linewidth, angle=0]{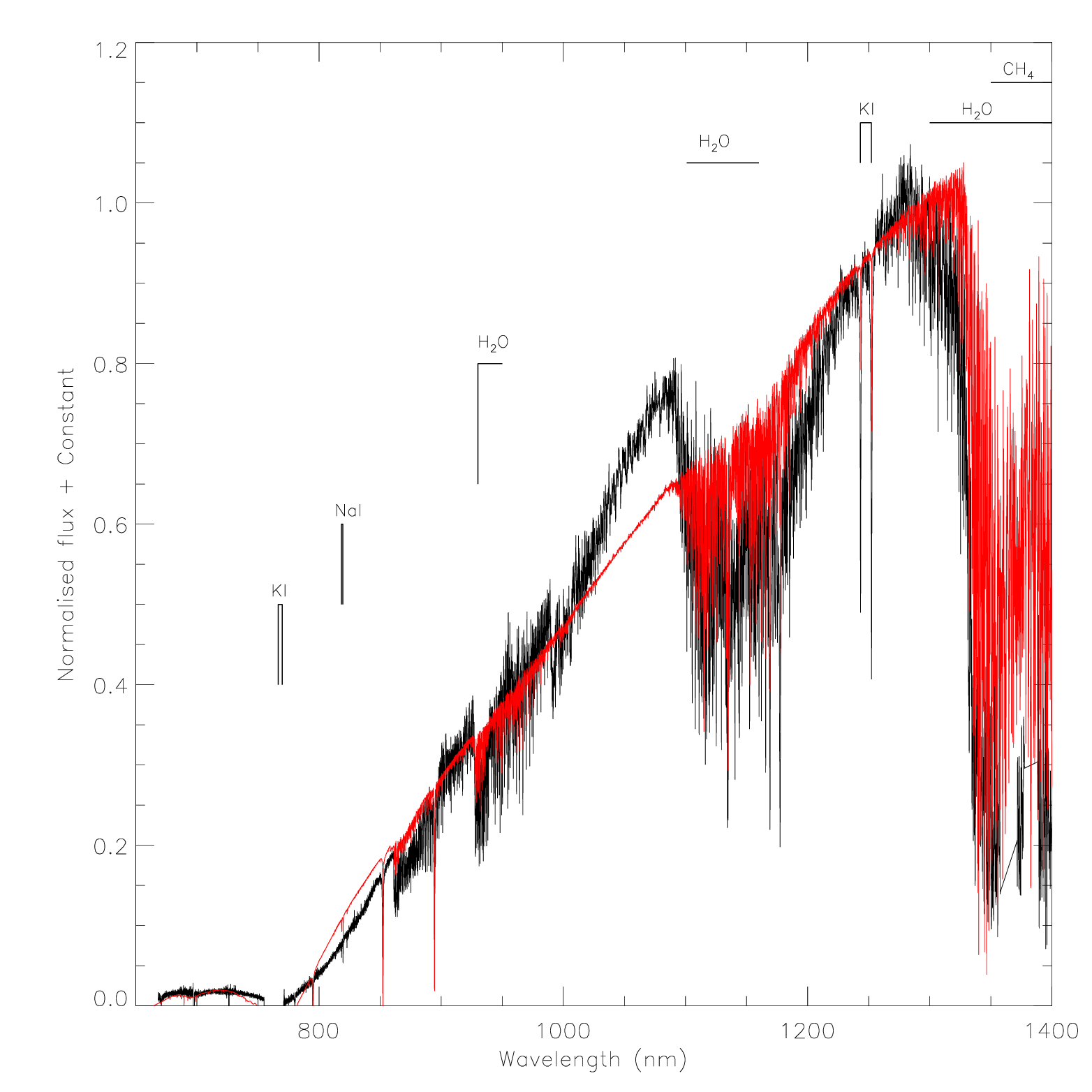}
  \includegraphics[width=0.49\linewidth, angle=0]{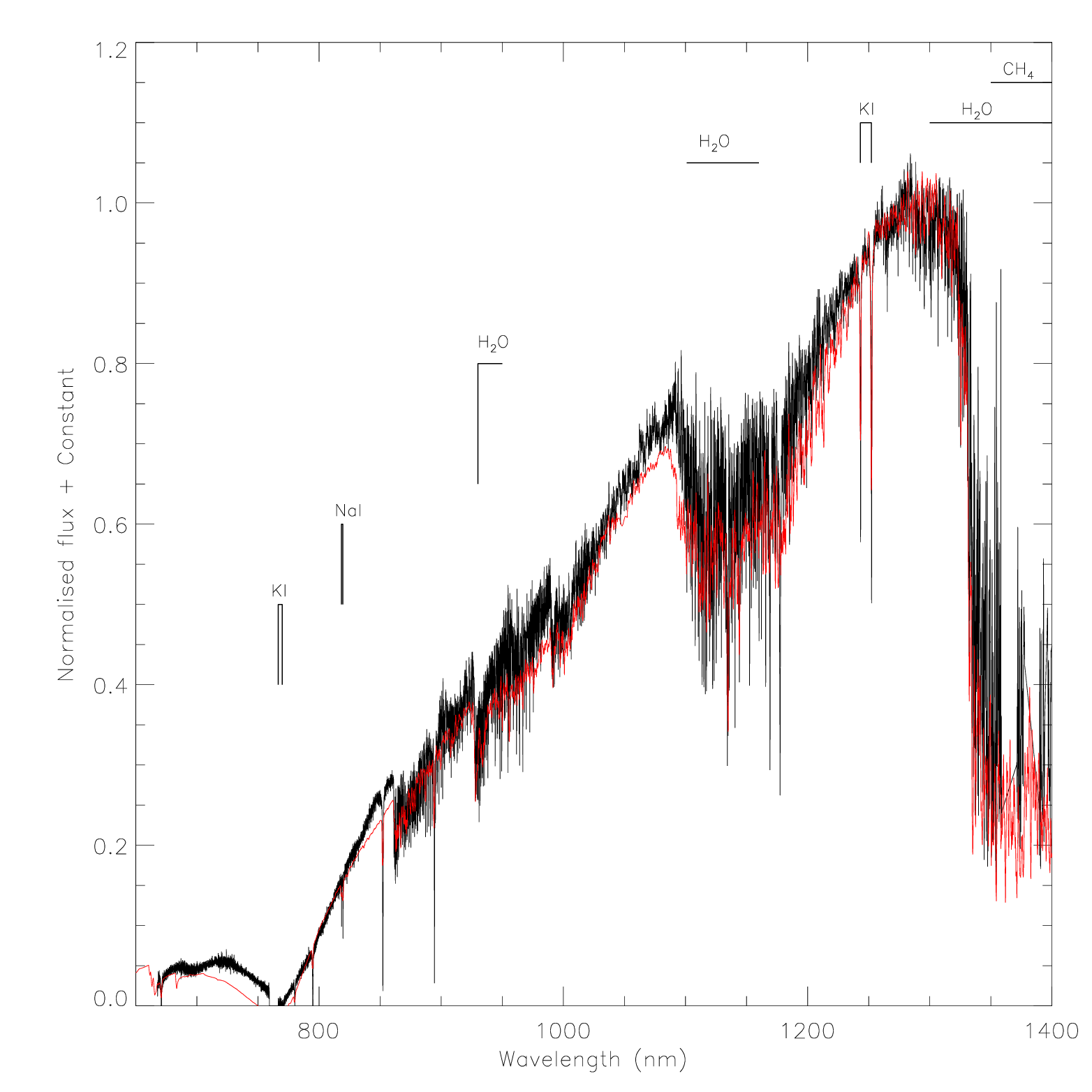}
  \includegraphics[width=0.49\linewidth, angle=0]{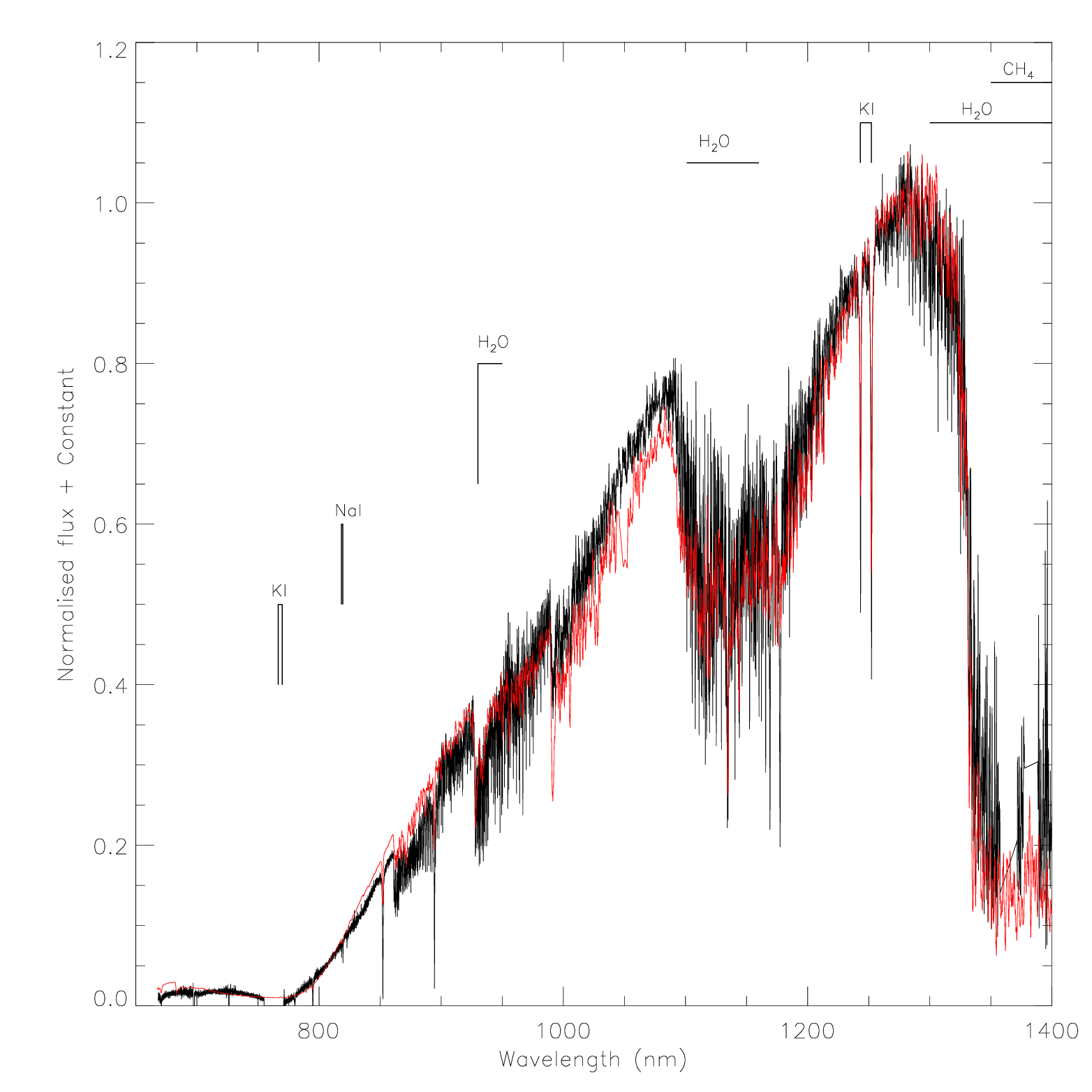}
   \caption{
{\it{Top:}} Comparison of the X-shooter optical and near-infrared
spectra (black line) of the L (left) and T (right) components of Luhman\,16AB
and the BT-Settl model atmosphere with 1400/5.0 (red line) and normalised
around 800 nm to match the potassium wings.
{\it{Bottom:}} Same X-shooter spectra (black line) compared to synthetic
spectra generated combining the WITA chemical code with the BT-Settl model
atmosphere for a temperature of 1400\,K, a gravity of $\log$(g) of 5.0--5.5, and
different relative abundances of Na, K, and other refractory elements
(smoothed for visualisation purposes).
   }
   \label{fig_Luhman16_XSH:spec_theory_NextGen}
\end{figure*}

We compared the observed spectrum of each component of the Luhman\,16AB system
with the BT-Settl synthetic models \citep{allard12}. We were unable to simultaneously 
reproduce the X-shooter data of the two objects with the same temperature, gravity and 
metallicity \citep[see also][]{faherty14a}. In Fig.\ \ref{fig_Luhman16_XSH:spec_theory_NextGen}, 
the BT-Settl spectrum computed for solar abundance, $\log$\,$g$\,=\,5.0 and $T_{\rm eff}$ = 1400 K 
(close to the values derived/discussed in previous sections) is shown in comparison with 
the X-shooter of Luhman\,16A and B\@.
However, according to \citet{faherty14a}, the BT-Settl models can reproduce each component 
of Luhman\,16 using different temperatures and surface gravities (i.e., ages), hypothesis unlikely 
for a nearby co-moving, lithium bearing, low-mass binary system. Using our code (see next), the fitting 
of the X-shooter spectra yields $T_{\rm eff}$\,=\,1500\,K and $\log$\,$g$\,=\,5.0 dex for the 
L component, and 1200--1300 K and 4.0 dex for the T component. This is differing temperatures and 
gravities, which agrees with the results of \citet{faherty14a}. As previously discussed, given the 
similar luminosities of the two pair members and assuming coevality, the $T_{\rm eff}$’s of the 
two objects cannot differ by more than $\sim$60 K (at the 2 $\sigma$ level). In the following, 
we intend to investigate whether one single temperature and gravity can provide reasonable fits 
to the observed data.

The observed differences between the X-shooter and computed BT-Settl spectra show the lack 
of our understanding of the physics of atmospheres of L and T dwarfs in general. To improve 
the fits with a special attention to the simultaneous fit of the sodium and potassium subordinate 
and resonance lines, we computed synthetic spectra of L and T dwarfs following the procedure by 
\citet{pavlenko07}, as summarised below.
We took advantage of the versatility of the WITA code to modify the relative abundances 
and number densities of species shaping the spectral energy distributions of L and T dwarfs.
We computed chemical equilibrium for a mix of approximately 100 molecular species and 
adopted the conventional approach of \citet{pavlenko13} to treat the depletion of atomic 
alkalis into molecular species. Then, we computed partial pressures of some molecules 
and atoms which exceeded the pressures of the gas-dust phase transitions and decreased 
their gas phase abundances to the corresponding equilibrium values \citep[see][]{pavlenko98a}.
We computed the profiles of the Na{\small{I}} and K{\small{I}} resonance doublets
in the framework of a quasi-static approach described in \citet{pavlenko07b} with an
upgraded approach from \citet{burrows03}. 
We computed the CrH and FeH line lists from the work of \citet{burrows02b} and
\citet{dulick03}, respectively, while the water vapour line list was taken from \citet{tennyson07}.
We considered the spectroscopic data for atomic absorption of the Vienna Atomic Line Database
\citep[VALD;][]{kupka99}\footnote{http://vald.astro.univie.ac.at/$\sim$vald/php/vald.php}.
We plot the overall spectral energy distribution produced by WITA after modifying
the number densities of Na{\small{I}}, K{\small{I}}, CrH, FeH, and water vapour for
L and T component in the bottom plots in Fig.\ \ref{fig_Luhman16_XSH:spec_theory_NextGen},
assuming an effective temperature of 1400\,K and a gravity of $\log$(g) of 5.0--5.5\@.
In that way we use rather phenomenological approach to fit our theoretical spectra to 
the observed spectral energy distributions.
The detailed analysis of these abundances deserve a more dedicated study which is beyond 
the scope of this paper. Nonetheless, it emerges that the spectral properties 
of L/T transition objects are not only determined by effective temperature and gravity but
also depend on the depletion of potassium and sodium atoms into condensates and dust species, 
as previously suggested by \citet{marley10}.

%
%
%

%
%
\section{Summary}
\label{Luhman16_XSH:conclusions}

We presented high-quality optical and near-infrared of the Luhman~16AB obtained
with VLT/X-shooter. We classify each component of this nearby brown dwarf binary
as L6--L7.5 and T0$\pm$1, respectively. We focussed our analysis primarily on the 
strength of alkali lines, which can be summarised as follows:
\begin{itemize}
\item[$\bullet$] We measured similar pEWs for the lithium absorption line in each component,
placing both objects in the substellar regime. 
The pEW measurement of the L component is consistent with the values of other late-L 
field dwarfs. Lithium is detected for the first time in a T dwarf (see also \citealt{faherty14a})
with an pEW of 8.4$\pm$1.5\AA{}
\item[$\bullet$] We set an upper limit of 30 m\AA{} on the lithium subordinate line at 812.64 nm.
\item[$\bullet$] The pEWs of the rubidium lines are similar for the L and T components of the
binary whereas the cesium lines appear stronger in the T dwarf.
\item[$\bullet$] We detected the sodium subordinate lines in the red optical spectra of the L and T components. 
The L-type member exhibits stronger sodium lines than the T component.
\item[$\bullet$] The pEWs of lithium and cesium decrease for spectral types later than $\sim$L6
and T0--T5, respectively, likely corresponding to the condensation of these species into
molecules in the atmospheres of cool brown dwarfs. Our comparison with other L and T
dwarfs shows that the condensation of cesium occurs at temperatures about 300\,K cooler 
than the one of lithium, as predicted by models.
\end{itemize}
We derived effective temperatures of 1305$^{+180}_{-135}$\,K and 1320\,$^{+185}_{-135}$\,K 
and luminosities of $-4.68 \pm 0.08$ dex and $-4.66 \pm 0.13$ dex for the L and T component 
of the binary, respectively, which are in agreement with the literature. 
Moreover, we measured heliocentric radial velocities of 25.1 and 19.3 km\,s$^{-1}$ 
for each component, yielding Galactic space velocities which do not indicate 
membership to any of the known, nearby, young moving group.
Finally, we fit synthetic spectra to the observed spectra of each component of the binary 
by combining the BT-Settl models and the WITA chemical abundance code. We reproduced 
satisfactorily the spectral energy distributions of the L and T dwarfs with a single set 
of temperature and gravity, but different relative atmospheric abundances of some elements 
like K, Na, Cr, which may be related to varying levels of dust condensation.

%
%
\begin{acknowledgements}
NL was funded by the Ram\'on y Cajal fellowship number 08-303-01-02\@.
This research has been supported by the Spanish Ministry of Economics and 
Competitiveness under the projects AYA2010-19136, AYA2010-21308-C3-02, 
AYA2010-21308-C03-03, AYA2010-20535, and AYA2011-30147-C03-03.\@.

This work is based on observations collected with X-shooter on the VLT
at the European Southern Observatory, Chile, under DDT programme 290.C-5200(B)
whose PI was Lodieu. We thank Jackie Faherty for sending us the optical
and near-infrared spectra of Luhman\,16AB published in her paper \citep{faherty14a}.
This research has made use of the Simbad and Vizier databases, operated
at the Centre de Donn\'ees Astronomiques de Strasbourg (CDS), and
of NASA's Astrophysics Data System Bibliographic Services (ADS).

This research has benefitted from the M, L, T, and Y dwarf compendium housed 
at DwarfArchives.org and the SpeX Prism Spectral Libraries, maintained by 
Adam Burgasser at http://pono.ucsd.edu/\~adam/browndwarfs/spexprism.

\end{acknowledgements}
%

%
%
\bibliographystyle{aa}
\bibliography{../../AA/mnemonic,../../AA/biblio_old}

\begin{thebibliography}{78}
\expandafter\ifx\csname natexlab\endcsname\relax\def\natexlab#1{#1}\fi

\bibitem[{{Allard} {et~al.}(2001){Allard}, {Hauschildt}, {Alexander},
  {Tamanai}, \& {Schweitzer}}]{allard01}
{Allard}, F., {Hauschildt}, P.~H., {Alexander}, D.~R., {Tamanai}, A., \&
  {Schweitzer}, A. 2001, ApJ, 556, 357

\bibitem[{{Allard} {et~al.}(2012){Allard}, {Homeier}, \& {Freytag}}]{allard12}
{Allard}, F., {Homeier}, D., \& {Freytag}, B. 2012, Royal Society of London
  Philosophical Transactions Series A, 370, 2765

\bibitem[{{Baraffe} {et~al.}(2003){Baraffe}, {Chabrier}, {Barman}, {Allard}, \&
  {Hauschildt}}]{baraffe03}
{Baraffe}, I., {Chabrier}, G., {Barman}, T.~S., {Allard}, F., \& {Hauschildt},
  P.~H. 2003, A\&A, 402, 701

\bibitem[{{Basri} {et~al.}(2000){Basri}, {Mohanty}, {Allard}, {Hauschildt},
  {Delfosse}, {Mart{\'{\i}}n}, {Forveille}, \& {Goldman}}]{basri00}
{Basri}, G., {Mohanty}, S., {Allard}, F., {et~al.} 2000, ApJ, 538, 363

\bibitem[{{Biller} {et~al.}(2013){Biller}, {Crossfield}, {Mancini}, {Ciceri},
  {Southworth}, {Kopytova}, {Bonnefoy}, {Deacon}, {Schlieder}, {Buenzli},
  {Brandner}, {Allard}, {Homeier}, {Freytag}, {Bailer-Jones}, {Greiner},
  {Henning}, \& {Goldman}}]{biller13b}
{Biller}, B.~A., {Crossfield}, I.~J.~M., {Mancini}, L., {et~al.} 2013, ApJL,
  778, L10

\bibitem[{{Boffin} {et~al.}(2014){Boffin}, {Pourbaix}, {Mu{\v z}i{\'c}},
  {Ivanov}, {Kurtev}, {Beletsky}, {Mehner}, {Berger}, {Girard}, \&
  {Mawet}}]{boffin14a}
{Boffin}, H.~M.~J., {Pourbaix}, D., {Mu{\v z}i{\'c}}, K., {et~al.} 2014, A\&A,
  561, L4

\bibitem[{{Burgasser} {et~al.}(2006){Burgasser}, {Geballe}, {Leggett},
  {Kirkpatrick}, \& {Golimowski}}]{burgasser06a}
{Burgasser}, A.~J., {Geballe}, T.~R., {Leggett}, S.~K., {Kirkpatrick}, J.~D.,
  \& {Golimowski}, D.~A. 2006, ApJ, 637, 1067

\bibitem[{{Burgasser} {et~al.}(2014){Burgasser}, {Gillon}, {Faherty},
  {Radigan}, {Triaud}, {Plavchan}, {Street}, {Jehin}, {Delrez}, \&
  {Opitom}}]{burgasser14a}
{Burgasser}, A.~J., {Gillon}, M., {Faherty}, J.~K., {et~al.} 2014, ApJ, 785, 48

\bibitem[{{Burgasser} {et~al.}(2003){Burgasser}, {Kirkpatrick}, {Liebert}, \&
  {Burrows}}]{burgasser03d}
{Burgasser}, A.~J., {Kirkpatrick}, J.~D., {Liebert}, J., \& {Burrows}, A. 2003,
  ApJ, 594, 510

\bibitem[{{Burgasser} {et~al.}(2004){Burgasser}, {McElwain}, {Kirkpatrick},
  {Cruz}, {Tinney}, \& {Reid}}]{burgasser04c}
{Burgasser}, A.~J., {McElwain}, M.~W., {Kirkpatrick}, J.~D., {et~al.} 2004, AJ,
  127, 2856

\bibitem[{{Burgasser} {et~al.}(2013){Burgasser}, {Sheppard}, \&
  {Luhman}}]{burgasser13b}
{Burgasser}, A.~J., {Sheppard}, S.~S., \& {Luhman}, K.~L. 2013, ArXiv e-prints

\bibitem[{{Burgasser} {et~al.}(2000){Burgasser}, {Wilson}, {Kirkpatrick},
  {Skrutskie}, {Colonno}, {Enos}, {Smith}, {Henderson}, {Gizis}, {Brown}, \&
  {Houck}}]{burgasser00b}
{Burgasser}, A.~J., {Wilson}, J.~C., {Kirkpatrick}, J.~D., {et~al.} 2000, AJ,
  120, 1100

\bibitem[{{Burrows} {et~al.}(2002{\natexlab{a}}){Burrows}, {Burgasser},
  {Kirkpatrick}, {Liebert}, {Milsom}, {Sudarsky}, \& {Hubeny}}]{burrows02a}
{Burrows}, A., {Burgasser}, A.~J., {Kirkpatrick}, J.~D., {et~al.}
  2002{\natexlab{a}}, \apj, 573, 394

\bibitem[{{Burrows} {et~al.}(2002{\natexlab{b}}){Burrows}, {Ram}, {Bernath},
  {Sharp}, \& {Milsom}}]{burrows02b}
{Burrows}, A., {Ram}, R.~S., {Bernath}, P., {Sharp}, C.~M., \& {Milsom}, J.~A.
  2002{\natexlab{b}}, ApJ, 577, 986

\bibitem[{{Burrows} \& {Sharp}(1999)}]{burrows99}
{Burrows}, A. \& {Sharp}, C.~M. 1999, ApJ, 512, 843

\bibitem[{{Burrows} {et~al.}(2003){Burrows}, {Sudarsky}, \&
  {Lunine}}]{burrows03}
{Burrows}, A., {Sudarsky}, D., \& {Lunine}, J.~I. 2003, ApJ, 596, 587

\bibitem[{{Chabrier} \& {Baraffe}(2000)}]{chabrier00a}
{Chabrier}, G. \& {Baraffe}, I. 2000, ARA\&A, 38, 337

\bibitem[{{Chiu} {et~al.}(2006){Chiu}, {Fan}, {Leggett}, {Golimowski}, {Zheng},
  {Geballe}, {Schneider}, \& {Brinkmann}}]{chiu06}
{Chiu}, K., {Fan}, X., {Leggett}, S.~K., {et~al.} 2006, AJ, 131, 2722

\bibitem[{{Crossfield} {et~al.}(2014){Crossfield}, {Biller}, {Schlieder},
  {Deacon}, {Bonnefoy}, {Homeier}, {Allard}, {Buenzli}, {Henning}, {Brandner},
  {Goldman}, \& {Kopytova}}]{crossfield14}
{Crossfield}, I.~J.~M., {Biller}, B., {Schlieder}, J.~E., {et~al.} 2014, Nat,
  505, 654

\bibitem[{{Cruz} {et~al.}(2009){Cruz}, {Kirkpatrick}, \& {Burgasser}}]{cruz09}
{Cruz}, K.~L., {Kirkpatrick}, J.~D., \& {Burgasser}, A.~J. 2009, AJ, 137, 3345

\bibitem[{{Cruz} {et~al.}(2003){Cruz}, {Reid}, {Liebert}, {Kirkpatrick}, \&
  {Lowrance}}]{cruz03}
{Cruz}, K.~L., {Reid}, I.~N., {Liebert}, J., {Kirkpatrick}, J.~D., \&
  {Lowrance}, P.~J. 2003, AJ, 126, 2421

\bibitem[{{Cushing} {et~al.}(2011){Cushing}, {Kirkpatrick}, {Gelino},
  {Griffith}, {Skrutskie}, {Mainzer}, {Marsh}, {Beichman}, {Burgasser},
  {Prato}, {Simcoe}, {Marley}, {Saumon}, {Freedman}, {Eisenhardt}, \&
  {Wright}}]{cushing11}
{Cushing}, M.~C., {Kirkpatrick}, J.~D., {Gelino}, C.~R., {et~al.} 2011, ApJ,
  743, 50

\bibitem[{{Dahn} {et~al.}(2002){Dahn}, {Harris}, {Vrba}, {Guetter}, {Canzian},
  {Henden}, {Levine}, {Luginbuhl}, {Monet}, {Monet}, {Pier}, {Stone}, {Walker},
  {Burgasser}, {Gizis}, {Kirkpatrick}, {Liebert}, \& {Reid}}]{dahn02}
{Dahn}, C.~C., {Harris}, H.~C., {Vrba}, F.~J., {et~al.} 2002, AJ, 124, 1170

\bibitem[{{D'Odorico} {et~al.}(2006){D'Odorico}, {Dekker}, {Mazzoleni},
  {Vernet}, {Guinouard}, {Groot}, {Hammer}, {Rasmussen}, {Kaper}, {Navarro},
  {Pallavicini}, {Peroux}, \& {Zerbi}}]{dOdorico06}
{D'Odorico}, S., {Dekker}, H., {Mazzoleni}, R., {et~al.} 2006, in Society of
  Photo-Optical Instrumentation Engineers (SPIE) Conference Series, Vol. 6269,
  Society of Photo-Optical Instrumentation Engineers (SPIE) Conference Series

\bibitem[{{Dulick} {et~al.}(2003){Dulick}, {Bauschlicher}, {Burrows}, {Sharp},
  {Ram}, \& {Bernath}}]{dulick03}
{Dulick}, M., {Bauschlicher}, Jr., C.~W., {Burrows}, A., {et~al.} 2003, ApJ,
  594, 651

\bibitem[{{Eggen}(1990)}]{eggen90a}
{Eggen}, O.~J. 1990, AJ, 100, 1159

\bibitem[{{Faherty} {et~al.}(2014){Faherty}, {Beletsky}, {Burgasser}, {Tinney},
  {Osip}, {Filippazzo}, \& {Simcoe}}]{faherty14a}
{Faherty}, J.~K., {Beletsky}, Y., {Burgasser}, A.~J., {et~al.} 2014, ApJ, 790,
  90

\bibitem[{{Geballe} {et~al.}(2002){Geballe}, {Knapp}, {Leggett}, {Fan},
  {Golimowski}, {Anderson}, {Brinkmann}, {Csabai}, \& {21
  coauthors}}]{geballe02}
{Geballe}, T.~R., {Knapp}, G.~R., {Leggett}, S.~K., {et~al.} 2002, ApJ, 564,
  466

\bibitem[{{Gillon} {et~al.}(2013){Gillon}, {Triaud}, {Jehin}, {Delrez},
  {Opitom}, {Magain}, {Lendl}, \& {Queloz}}]{gillon13a}
{Gillon}, M., {Triaud}, A.~H.~M.~J., {Jehin}, E., {et~al.} 2013, A\&A, 555, L5

\bibitem[{{Golimowski} {et~al.}(2004{\natexlab{a}}){Golimowski}, {Henry},
  {Krist}, {Dieterich}, {Ford}, {Illingworth}, {Ardila}, {Clampin}, {Franz},
  {Wasserman}, {Benedict}, {McArthur}, \& {Nelan}}]{golimowski04b}
{Golimowski}, D.~A., {Henry}, T.~J., {Krist}, J.~E., {et~al.}
  2004{\natexlab{a}}, AJ, 128, 1733

\bibitem[{{Golimowski} {et~al.}(2004{\natexlab{b}}){Golimowski}, {Leggett},
  {Marley}, {Fan}, {Geballe}, {Knapp}, {Vrba}, {Henden}, \& {11
  authors}}]{golimowski04a}
{Golimowski}, D.~A., {Leggett}, S.~K., {Marley}, M.~S., {et~al.}
  2004{\natexlab{b}}, AJ, 127, 3516

\bibitem[{{Johnson} \& {Soderblom}(1987)}]{johnson87}
{Johnson}, D.~R.~H. \& {Soderblom}, D.~R. 1987, AJ, 93, 864

\bibitem[{{King} {et~al.}(2010){King}, {McCaughrean}, {Homeier}, {Allard},
  {Scholz}, \& {Lodieu}}]{king10b}
{King}, R.~R., {McCaughrean}, M.~J., {Homeier}, D., {et~al.} 2010, A\&A, 510,
  A99

\bibitem[{{Kirkpatrick} {et~al.}(2008){Kirkpatrick}, {Cruz}, {Barman},
  {Burgasser}, {Looper}, {Tinney}, {Gelino}, {Lowrance}, {Liebert},
  {Carpenter}, {Hillenbrand}, \& {Stauffer}}]{kirkpatrick08}
{Kirkpatrick}, J.~D., {Cruz}, K.~L., {Barman}, T.~S., {et~al.} 2008, \apj, 689,
  1295

\bibitem[{{Kirkpatrick} {et~al.}(2012){Kirkpatrick}, {Gelino}, {Cushing},
  {Mace}, {Griffith}, {Skrutskie}, {Marsh}, {Wright}, {Eisenhardt}, {McLean},
  {Mainzer}, {Burgasser}, {Tinney}, {Parker}, \& {Salter}}]{kirkpatrick12}
{Kirkpatrick}, J.~D., {Gelino}, C.~R., {Cushing}, M.~C., {et~al.} 2012, ApJ,
  753, 156

\bibitem[{{Kirkpatrick} {et~al.}(1999){Kirkpatrick}, {Reid}, {Liebert},
  {Cutri}, {Nelson}, {Beichman}, {Dahn}, {Monet}, {Gizis}, \&
  {Skrutskie}}]{kirkpatrick99}
{Kirkpatrick}, J.~D., {Reid}, I.~N., {Liebert}, J., {et~al.} 1999, ApJ, 519,
  802

\bibitem[{{Kirkpatrick} {et~al.}(2000){Kirkpatrick}, {Reid}, {Liebert},
  {Gizis}, {Burgasser}, {Monet}, {Dahn}, {Nelson}, \&
  {Williams}}]{kirkpatrick00}
{Kirkpatrick}, J.~D., {Reid}, I.~N., {Liebert}, J., {et~al.} 2000, AJ, 120, 447

\bibitem[{{Knapp} {et~al.}(2004){Knapp}, {Leggett}, {Fan}, {Marley}, {Geballe},
  {Golimowski}, {Finkbeiner}, {Gunn}, , \& {21 co-authors}}]{knapp04}
{Knapp}, G.~R., {Leggett}, S.~K., {Fan}, X., {et~al.} 2004, AJ, 127, 3553

\bibitem[{{Kniazev} {et~al.}(2013){Kniazev}, {Vaisanen}, {Mu{\v z}i{\'c}},
  {Mehner}, {Boffin}, {Kurtev}, {Melo}, {Ivanov}, {Girard}, {Mawet},
  {Schmidtobreick}, {Huelamo}, {Borissova}, {Minniti}, {Ishibashi}, {Potter},
  {Beletsky}, {Buckley}, {Crawford}, {Gulbis}, {Kotze}, {Miszalski},
  {Pickering}, {Romero Colmenero}, \& {Williams}}]{kniazev13}
{Kniazev}, A.~Y., {Vaisanen}, P., {Mu{\v z}i{\'c}}, K., {et~al.} 2013, ApJ,
  770, 124

\bibitem[{{Kupka} {et~al.}(1999){Kupka}, {Piskunov}, {Ryabchikova}, {Stempels},
  \& {Weiss}}]{kupka99}
{Kupka}, F., {Piskunov}, N., {Ryabchikova}, T.~A., {Stempels}, H.~C., \&
  {Weiss}, W.~W. 1999, A\&AS, 138, 119

\bibitem[{{Leggett}(1992)}]{leggett92}
{Leggett}, S.~K. 1992, ApJS, 82, 351

\bibitem[{{Leggett} {et~al.}(2000){Leggett}, {Geballe}, {Fan}, {Schneider},
  {Gunn}, {Lupton}, {Knapp}, {Strauss}, \& {27 coauthors}}]{leggett00}
{Leggett}, S.~K., {Geballe}, T.~R., {Fan}, X., {et~al.} 2000, ApJL, 536, L35

\bibitem[{{Leggett} {et~al.}(2002){Leggett}, {Golimowski}, {Fan}, {Geballe},
  {Knapp}, {Brinkmann}, {Csabai}, {Gunn}, \& {11 co-authors}}]{leggett02}
{Leggett}, S.~K., {Golimowski}, D.~A., {Fan}, X., {et~al.} 2002, ApJ, 564, 452

\bibitem[{{Lemoine} {et~al.}(1993){Lemoine}, {Ferlet}, {Vidal-Madjar},
  {Emerich}, \& {Bertin}}]{lemoine93}
{Lemoine}, M., {Ferlet}, R., {Vidal-Madjar}, A., {Emerich}, C., \& {Bertin}, P.
  1993, A\&A, 269, 469

\bibitem[{{Lodders} \& {Fegley}(2006)}]{lodders06}
{Lodders}, K. \& {Fegley}, Jr., B. 2006, {Chemistry of Low Mass Substellar
  Objects}, ed. J.~W. {Mason}, 1

\bibitem[{{Looper} {et~al.}(2007){Looper}, {Kirkpatrick}, \&
  {Burgasser}}]{looper07}
{Looper}, D.~L., {Kirkpatrick}, J.~D., \& {Burgasser}, A.~J. 2007, AJ, 134,
  1162

\bibitem[{{Luhman}(2013)}]{luhman13a}
{Luhman}, K.~L. 2013, ApJL, 767, L1

\bibitem[{{Mamajek}(2013)}]{mamajek13a}
{Mamajek}, E.~E. 2013, ArXiv e-prints

\bibitem[{{Marley} {et~al.}(2010){Marley}, {Saumon}, \& {Goldblatt}}]{marley10}
{Marley}, M.~S., {Saumon}, D., \& {Goldblatt}, C. 2010, ApJL, 723, L117

\bibitem[{{Mart\'{\i}n} {et~al.}(1999){Mart\'{\i}n}, {Delfosse}, {Basri},
  {Goldman}, {Forveille}, \& {Zapatero Osorio}}]{martin99a}
{Mart\'{\i}n}, E.~L., {Delfosse}, X., {Basri}, G., {et~al.} 1999, AJ, 118, 2466

\bibitem[{{McCaughrean} {et~al.}(2004){McCaughrean}, {Close}, {Scholz},
  {Lenzen}, {Biller}, {Brandner}, {Hartung}, \& {Lodieu}}]{mjm04}
{McCaughrean}, M.~J., {Close}, L.~M., {Scholz}, R.-D., {et~al.} 2004, A\&A,
  413, 1029

\bibitem[{{Nakajima} {et~al.}(1995){Nakajima}, {Oppenheimer}, {Kulkarni},
  {Golimowski}, {Matthews}, \& {Durrance}}]{nakajima95}
{Nakajima}, T., {Oppenheimer}, B.~R., {Kulkarni}, S.~R., {et~al.} 1995, Nat,
  378, 463

\bibitem[{{Nakajima} {et~al.}(2004){Nakajima}, {Tsuji}, \&
  {Yanagisawa}}]{nakajima04a}
{Nakajima}, T., {Tsuji}, T., \& {Yanagisawa}, K. 2004, ApJ, 607, 499

\bibitem[{{Nelson} {et~al.}(1993){Nelson}, {Rappaport}, \&
  {Chiang}}]{nelson93b}
{Nelson}, L.~A., {Rappaport}, S., \& {Chiang}, E. 1993, ApJ, 413, 364

\bibitem[{{Pavlenko} {et~al.}(2000){Pavlenko}, {Zapatero Osorio}, \&
  {Rebolo}}]{pavlenko00a}
{Pavlenko}, Y., {Zapatero Osorio}, M.~R., \& {Rebolo}, R. 2000, A\&A, 355, 245

\bibitem[{{Pavlenko}(1998)}]{pavlenko98a}
{Pavlenko}, Y.~V. 1998, Astronomy Reports, 42, 787

\bibitem[{{Pavlenko}(2013)}]{pavlenko13}
{Pavlenko}, Y.~V. 2013, 84, 1062

\bibitem[{{Pavlenko} {et~al.}(2007{\natexlab{a}}){Pavlenko}, {Jones},
  {Mart{\'{\i}}n}, {Guenther}, {Kenworthy}, \& {Zapatero Osorio}}]{pavlenko07}
{Pavlenko}, Y.~V., {Jones}, H.~R.~A., {Mart{\'{\i}}n}, E.~L., {et~al.}
  2007{\natexlab{a}}, MNRAS, 380, 1285

\bibitem[{{Pavlenko} {et~al.}(2007{\natexlab{b}}){Pavlenko}, {Zhukovska}, \&
  {Volobuev}}]{pavlenko07b}
{Pavlenko}, Y.~V., {Zhukovska}, S.~V., \& {Volobuev}, M. 2007{\natexlab{b}},
  Astronomy Reports, 51, 282

\bibitem[{{Rebolo} {et~al.}(1992){Rebolo}, {Mart\'{\i}n}, \& {Magazz{\`
  u}}}]{rebolo92}
{Rebolo}, R., {Mart\'{\i}n}, E.~L., \& {Magazz{\` u}}, A. 1992, ApJL, 389, L83

\bibitem[{{Rebolo} {et~al.}(1995){Rebolo}, {Zapatero-Osorio}, \&
  {Mart\'{\i}n}}]{rebolo95}
{Rebolo}, R., {Zapatero-Osorio}, M.~R., \& {Mart\'{\i}n}, E.~L. 1995, Nat, 377,
  129

\bibitem[{{Reid} {et~al.}(2001){Reid}, {Burgasser}, {Cruz}, {Kirkpatrick}, \&
  {Gizis}}]{reid01a}
{Reid}, I.~N., {Burgasser}, A.~J., {Cruz}, K.~L., {Kirkpatrick}, J.~D., \&
  {Gizis}, J.~E. 2001, AJ, 121, 1710

\bibitem[{{Reid} {et~al.}(2000){Reid}, {Kirkpatrick}, {Gizis}, {Dahn}, {Monet},
  {Williams}, {Liebert}, \& {Burgasser}}]{reid00}
{Reid}, I.~N., {Kirkpatrick}, J.~D., {Gizis}, J.~E., {et~al.} 2000, AJ, 119,
  369

\bibitem[{{Scholz} {et~al.}(2003){Scholz}, {McCaughrean}, {Lodieu}, \&
  {Kuhlbrodt}}]{scholz03}
{Scholz}, R.-D., {McCaughrean}, M.~J., {Lodieu}, N., \& {Kuhlbrodt}, B. 2003,
  A\&A, 398, L29

\bibitem[{{Stephens} {et~al.}(2009){Stephens}, {Leggett}, {Cushing}, {Marley},
  {Saumon}, {Geballe}, {Golimowski}, {Fan}, \& {Noll}}]{stephens09}
{Stephens}, D.~C., {Leggett}, S.~K., {Cushing}, M.~C., {et~al.} 2009, ApJ, 702,
  154

\bibitem[{{Tennyson} {et~al.}(2007){Tennyson}, {Harris}, {Barber}, {La Delfa},
  {Voronin}, {Kaminsky}, \& {Pavlenko}}]{tennyson07}
{Tennyson}, J., {Harris}, G.~J., {Barber}, R.~J., {et~al.} 2007, Molecular
  Physics, 105, 701

\bibitem[{{Tinney}(1998)}]{tinney98a}
{Tinney}, C.~G. 1998, MNRAS, 296, L42

\bibitem[{{Tinney} {et~al.}(2012){Tinney}, {Faherty}, {Kirkpatrick}, {Wright},
  {Gelino}, {Cushing}, {Griffith}, \& {Salter}}]{tinney12}
{Tinney}, C.~G., {Faherty}, J.~K., {Kirkpatrick}, J.~D., {et~al.} 2012, ApJ,
  759, 60

\bibitem[{{Tody}(1986)}]{tody86}
{Tody}, D. 1986, in Society of Photo-Optical Instrumentation Engineers (SPIE)
  Conference Series, Vol. 627, Society of Photo-Optical Instrumentation
  Engineers (SPIE) Conference Series, ed. D.~L. {Crawford}, 733

\bibitem[{{Tody}(1993)}]{tody93}
{Tody}, D. 1993, in Astronomical Society of the Pacific Conference Series,
  Vol.~52, Astronomical Data Analysis Software and Systems II, ed. R.~J.
  {Hanisch}, R.~J.~V. {Brissenden}, \& J.~{Barnes}, 173

\bibitem[{{Torres} {et~al.}(2006){Torres}, {Quast}, {da Silva}, {de La Reza},
  {Melo}, \& {Sterzik}}]{torres06}
{Torres}, C.~A.~O., {Quast}, G.~R., {da Silva}, L., {et~al.} 2006, A\&A, 460,
  695

\bibitem[{{Torres} {et~al.}(2008){Torres}, {Quast}, {Melo}, \&
  {Sterzik}}]{torres08}
{Torres}, C.~A.~O., {Quast}, G.~R., {Melo}, C.~H.~F., \& {Sterzik}, M.~F. 2008,
  {Young Nearby Loose Associations}, ed. B.~{Reipurth}, 757

\bibitem[{{van Leeuwen}(2007)}]{vanLeeuwen07}
{van Leeuwen}, F. 2007, A\&A, 474, 653

\bibitem[{{Vernet} {et~al.}(2011){Vernet}, {Dekker}, {D'Odorico}, {Kaper},
  {Kjaergaard}, {Hammer}, {Randich}, {Zerbi}, \& {82 co-authors}}]{vernet11}
{Vernet}, J., {Dekker}, H., {D'Odorico}, S., {et~al.} 2011, A\&A, 536, A105

\bibitem[{{Vrba} {et~al.}(2004){Vrba}, {Henden}, {Luginbuhl}, {Guetter},
  {Munn}, {Canzian}, {Burgasser}, {Kirkpatrick}, {Fan}, {Geballe},
  {Golimowski}, {Knapp}, {Leggett}, {Schneider}, \& {Brinkmann}}]{vrba04}
{Vrba}, F.~J., {Henden}, A.~A., {Luginbuhl}, C.~B., {et~al.} 2004, AJ, 127,
  2948

\bibitem[{{Wright} {et~al.}(2010){Wright}, {Eisenhardt}, {Mainzer}, {Ressler},
  {Cutri}, {Jarrett}, {Kirkpatrick}, \& {31 co-authors}}]{wright10}
{Wright}, E.~L., {Eisenhardt}, P.~R.~M., {Mainzer}, A.~K., {et~al.} 2010, AJ,
  140, 1868

\bibitem[{{Zapatero Osorio} {et~al.}(2014){Zapatero Osorio}, {B{\'e}jar},
  {Miles-P{\'a}ez}, {Pe{\~n}a Ram{\'{\i}}rez}, {Rebolo}, \&
  {Pall{\'e}}}]{zapatero14a}
{Zapatero Osorio}, M.~R., {B{\'e}jar}, V.~J.~S., {Miles-P{\'a}ez}, P.~A.,
  {et~al.} 2014, A\&A, 568, A6

\bibitem[{{Zuckerman} \& {Song}(2004)}]{zuckerman04}
{Zuckerman}, B. \& {Song}, I. 2004, ARA\&A, 42, 685

\end{thebibliography}
%

\end{document}